\documentclass[aps,prb,amssymb,showpacs,superscriptaddress,twocolumn,
tightenlines]{revtex4}%
\usepackage{graphicx}
\usepackage{amsmath,bm}
\usepackage[mathscr]{euscript}
\usepackage{dcolumn}

\begin{document}
\title{Theory for Reliable First-Principles Prediction of the Superconducting 
Transition Temperature}
\author{Yasutami Takada}
\thanks{Email: takada@issp.u-tokyo.ac.jp; published in ``Carbon-based 
superconductors: Toward high-$T_c$ superconductivity", edited by J. Haruyama 
(Pan Stanford, Singapore,2015), pp. 193-230; ISBN: 978-981-4303-30-9 (Hardcover), 
978-981-4303-31-6 (eBook).}
\affiliation{Institute for Solid State Physics, University of Tokyo,
             5-1-5 Kashiwanoha, Kashiwa, Chiba 277-8581, Japan}
\begin{abstract}
A review is given for the theoretical framework to give a reliable prediction 
of the superconducting transition temperature $T_c$ from first principles, 
together with a practical strategy for its application to actual materials 
with illustrations of the results of $T_c$ calculated for superconductors 
in the weak-coupling region like the alkali and alkaline-earth intercalated 
graphites as well as those in the strong-coupling region like the alkali-doped 
fullerides. 
\end{abstract}


\maketitle

\section{Introduction}
\label{sec_takada.1}

In quantum mechanics, a ground state is determined through a compromise between 
the kinetic energy (which makes particles itinerant) and the potential energy 
(which makes them localized). If the latter includes the interaction between 
particles, there appears a further complication due to their correlated motion. 
In elucidating the microscopic mechanism of superconductivity, this intrinsic 
complexity in quantum mechanics cannot be avoided but is even more intensified, 
specifically because superconductivity is a phenomenon in which an assembly of 
electrons, negatively charged particles with one-half spin, goes into the 
pair-condensed phase as a consequence of the dominance of some effective 
attractions between electrons mediated by either phonons, plasmons, 
spin-fluctuations, or orbital-fluctuations over the short-range Coulomb 
repulsions, indicating the necessity of deeply understanding and carefully 
investigating the physics of this charge-spin-phonon(-orbital) complex 
before making a reliable evaluation of the transition temperature $T_c$ of 
this second-order phase transition. Thus one would imagine that the task of 
reliably calculating $T_c$ must be formidably difficult, but the ultimate goal 
in the theoretical study of high-$T_c$ superconductivity should be to construct 
a good theoretical framework for an accurate prediction of $T_c$; without such 
a theoretical tool, we could never conduct a research directly and intimately 
touched with the most salient feature of high-$T_c$ materials, namely, the very 
feature that $T_c$ becomes very high in those materials. 

McMillan was the first to provide a rather successful scheme for 
predicting $T_c$ in the phonon mechanism of superconductivity, starting from 
the microscopic electron-phonon coupled Hamiltonian. The scheme is known as 
the McMillan's formula~\cite{McMillan68}, which was revised later by Allen and 
Dynes~\cite{Allen75,Allen82,Carbotte90}. The formulae, both original and 
revised, are derived from the Eliashberg theory of 
superconductivity~\cite{Eliashberg60} and the task of a microscopic calculation 
of $T_c$ in this framework is reduced to the evaluation of the so-called 
Eliashberg function $\alpha^2F(\omega)$ from the first-principles Hamiltonian, 
where $F(\omega)$ is the phonon density of states which may be observed by 
neutron diffraction. This function $\alpha^2F(\omega)$ enables us to obtain 
both the nondimensional electron-phonon coupling constant $\lambda$ and the 
average phonon energy $\omega_0$, through which we can give a first-principles 
prediction of $T_c$ with an additional introduction of a phenomenological 
parameter $\mu^*$ (the Coulomb pseudopotential~\cite{Morel62}) for the purpose 
of roughly estimating the effect of the short-range Coulomb repulsion between 
electrons on $T_c$. 

At present, this framework is usually regarded as the standard one for making 
a first-principles prediction of $T_c$ and widely used. In fact, the 
superconducting mechanism of many (so-called weakly-correlated) superconductors 
is believed to be clarified by employing this scheme. The key phonon modes to 
bring about superconductivity are identified by investigating the structure of 
$\alpha^2F(\omega)$. We can mention that superconductivity in MgB$_2$ with 
$T_c=39$K provides a very good example~\cite{Bohnen01,Kong01,Choi02a,Choi02b} 
to illustrate the power of this scheme. The case of CaC$_6$ with $T_c=11.5$K 
seems to constitute another recent example~\cite{Mazin,Mauri}. 

In spite of these and many other successful examples, however, this is not 
considered to be our ultimate scheme for calculating $T_c$ from first principles, 
primarily because a phenomenological parameter $\mu^*$ is included in the theory. 
Actually, it cannot be regarded as the method of predicting $T_c$ in the true 
sense of the word, if the parameter $\mu^*$ is determined so as to reproduce the 
observed $T_c$. Besides, as long as $\mu^*$ is employed to avoid a serious 
investigation of the effects of the Coulomb repulsion on superconductivity, this 
scheme cannot be applied to strongly-correlated superconductors such as the 
high-$T_c$ cuprates. Even in weakly- or moderately-correlated superconductors, 
this scheme cannot treat superconductivity originating from the Coulomb repulsion 
via charge, spin, and/or orbital fluctuations (namely, the electronic mechanism 
including the plasmon mechanism~\cite{Takada78,Takada93a}). Furthermore, in 
this scheme, we cannot investigate the competition or the coexistence (or even 
the mutual enhancement due to the quantum-mechanical constructive interference 
effect) between the phonon and the electronic mechanisms. 

The validity of the concept of $\mu^*$ is closely related to that of the 
Eliashberg theory itself; the theory is valid only if the Fermi energy 
of the superconducting electronic system, $E_{\rm F}$, is much larger 
than $\omega_0$. Note that under the condition of $E_{\rm F}\gg \omega_0$, the 
dynamical response time for the phonon-mediated attraction $\omega_0^{-1}$ is 
much slower than that for the Coulomb repulsion $E_{\rm F}^{-1}$, precluding 
any possible interference effects between two interactions, so that physically 
it is very plausible to separate them. After this separation, the Coulomb part 
(which was not anticipated to play a positive role in the Cooper-pair formation) 
has been simply treated in terms of a single parameter $\mu^*$. Thus, for the 
purpose of searching for some positive role of the Coulomb repulsion in 
superconductivity, the concept of $\mu^*$ is irrelevant from the outset of the 
whole theory. 

As for the condition of $E_{\rm F}\gg \omega_0$, it must also be noted that 
such a condition is violated in some recently discovered superconductors in 
the phonon mechanism including the alkali-doped fullerenes with $T_c = 
18-38$K~\cite{Hebard91,Takabayashi09,Gunnarsson97,Takada98}. Once it is 
violated, we need to include higher-order corrections in the electron-phonon 
coupling (or the so-called vertex corrections $\Gamma$) in calculating the 
phonon-mediated attractive interaction~\cite{Takada93b}. Then, it is by no 
means clear whether we can fully treat the overall effect of various phonons 
in terms of the sum of the contribution from each phonon. This implies that 
the Eliashberg function $\alpha^2F(\omega)$ will not be appropriate enough to 
describe the phonon-mediated attraction because of possible interference 
effects among virtually-excited different phonon modes. As a consequence, 
$\lambda$ will not be simply the sum of $\lambda_i$ the contribution from 
the $i$th phonon, unless $\lambda_i$ is small enough to validate the whole 
calculation in lowest-order perturbation. 

If the condition of $E_{\rm F}\gg \omega_0$ is violated, especially if 
$E_{\rm F}$ is about the same as $\omega_0$, another complication occurs in 
treating the screening effect of the conduction electrons. In the usual 
calculation scheme from first principles, the static screening is assumed in 
calculating $\alpha^2F(\omega)$, but it does not reflect the actual screening 
process working during the formation of Cooper pairs. This subtle problem of 
screening is, of course, also closely related to the problem of the 
first-principles determination of $\mu^*$ and we will not be able to solve these 
problems unambiguously without confronting with a difficult task of treating 
both the Coulomb repulsion and the phonon-mediated attraction on the same 
footing in the calculation of the microscopic {\it dynamical} electron-electron 
effective interaction $V$. 

In order to overcome the above-mentioned problems inherently associated with 
the Eliashberg theory, the first and natural option would be to improve on it 
by considering both the gap equation and the electron-electron effective 
interaction $V$ in entire energy- and momentum-space with properly including 
the vertex corrections $\Gamma$ in $V$ and without separating the Coulomb 
repulsion from $V$. However, this will not be easily accomplished at least in 
the near future, partly because the demand for computational resources becomes 
too much in the solution of the full {\it nonlocal} and {\it dynamical} gap 
equation and partly because no controlled approximation scheme has been known 
for $\Gamma$ for the superconducting state. (Note that the controlled scheme 
is known for the normal state~\cite{Takada95,Takada01}.)

Fortunately, an alternative option has already been proposed by the extension 
of the density functional theory (DFT) to treating a superconducting 
state~\cite{Oliveira88,Kurth99}. This theory provides a formally {\it exact} 
framework for calcualting $T_c$ from first principles by the solution of the gap 
equation only in momentum-space, setting aside the calculation of other physical 
quantities except for the one-electron density $n({\bm r})$. Note that the 
effect of the Coulomb repulsion is properly included in this formulation without 
resort to the concept of $\mu^*$. Therefore we shall begin with making a very 
brief review of this density fuctional theory for superconductors (SCDFT) in 
Sec.~\ref{sec_takada.2}. We shall point out that the central quantity in this 
framework is the pairing interaction ${\cal K}$. Then in Sec.~\ref{sec_takada.3}, 
we shall infer a concrete formula for ${\cal K}$ defined in terms of the 
Kohn-Sham orbitals in an inhomogeneous electron gas by reconsidering the gap 
equation for the homogeneous electron gas in the weak-coupling region with use 
of the Green's-function method. The formula has not been proposed so far in the 
literature in SCDFT, but by its application to superconductivity in the alkali- 
and alkaline-earth intercalated graphites~\cite{Takada09a,Takada09b}, it turns 
out that this is indeed a very good approximate functional form for ${\cal K}$. 
A prediction for the optimum value of $T_c$ by using this functional form will 
be given for this class of materials. In Sec.~\ref{sec_takada.4}, we shall 
consider a formula for ${\cal K}$ in the opposite limit, namely, in the 
strong-coupling region, especially for such superconductors with short coherence 
lengths as the alkali-doped fullerides~\cite{Takada07}. By interpolating the 
formulae for ${\cal K}$ in these two limits, we shall propose a new functional 
form for ${\cal K}$ which is supposed to work well in the whole range of the 
coupling strength. The results of $T_c$ obtained by its application to the 
fullerene superconductors and related materials will be shown in the last 
subsection of this section. Finally in Sec.~\ref{sec_takada.5} we shall conclude 
this short article, together with discussing the direction of future research. 

\section{Density Functional Theory for Superconductors (SCDFT)}
\label{sec_takada.2}

\subsection{Hohenberg-Kohn-Sham Theorem}

According to the basic theorem in the density functional theory (DFT) due to 
Hohenberg and Kohn~\cite{Hohenberg64}, all the physical quantities of an 
interacting electron system are uniquely determined, once its electronic density 
in the ground state $n({\bm r})$ is specified. This implies that every quantity 
including the exchange-correlation energy $F_{xc}$ may be considered as a unique 
functional of $n({\bm r})$. The ground-state density $n({\bm r})$ itself can be 
determined by the calculation of the ground-state electronic density of the 
corresponding noninteracting reference system that is stipulated in terms of 
the Kohn-Sham (KS) equation~\cite{Kohn65}. The concept of the noninteracting 
reference system is of central importance in the KS algorithm and the core 
quantity in the KS equation is the exchange-correlation potential 
$V_{xc}({\bm r})$, which is formally defined as the first-order functional 
derivative of $F_{xc}[n({\bm r})]$ with respect to $n({\bm r})$, namely, 
$V_{xc}({\bm r})=\delta F_{xc}[n]/\delta n({\bm r})$. It must be noted that 
$V_{xc}({\bm r})$ as well as each one-electronic wavefunction at $i$th level 
(usually called as ``$i$th KS orbital'') with its energy eigenvalue 
$\varepsilon_i$ in the KS equation has no direct physical relevance; they are 
merely introduced for the mathematical convenience so as to obtain the 
{\it exact} $n({\bm r})$ in the real many-electron system by exploitation of 
its one-to-one correspondence to the noninteracting reference system. 

This basic Hohenberg-Kohn theorem can be applied not only to the normal ground 
state but also to the ordered one on the understanding that the order parameter 
itself in the ordered state is regarded as a functional of $n({\bm r})$. 
In providing some approximate functional form for $F_{xc}[n]$ in actual 
calculations, however, it would be more convenient to treat the order parameter 
as an additional independent variable. For example, in considering the system 
with a collinear magnetic order, we usually employ the spin-dependent scheme in 
which the fundamental variable is not $n({\bm r})$ but the spin-decomposed 
density $n_{\sigma}({\bm r})$, leading to the spin-polarized exchange-correlation 
energy functional $F_{xc}[n_{\sigma}]$, based on which the spin-dependent 
exchange-correlation potential is defined to specify the spin-dependent KS 
equation for determining $n_{\sigma}({\bm r})$ from first principles. 

\subsection{Gap Equation in SCDFT}

In an essentially similar way, in treating superconductivity in the framework 
of DFT, it would be better to construct the energy functional with employing 
both $n({\bm r})$ and the electron-pair density (or the superconducting order 
parameter) $\chi({\bm r},{\bm r'}) (\equiv \langle \Psi_{\uparrow}({\bm r})
\Psi_{\downarrow}({\bm r'})\rangle)$ as basic variables~\cite{Oliveira88,Kurth99}, 
leading to the pair-density-dependent exchange-correlation energy functional 
$F_{xc}[n ({\bm r}),\chi({\bm r,r'})]$, where $\Psi_{\sigma}({\bm r})$ is the 
annihilation operator of $\sigma$-spin electron field at position ${\bm r}$. 
In accordance with this addition of the order parameter as a fundamental 
variable to DFT, not only the exchange-correlation potential $V_{xc}({\bm r})$ 
but also the exchange-correlation pair-potential $\Delta_{xc}({\bm r},{\bm r'})
=-\delta F_{xc}[n,\chi]/\delta \chi^*({\bm r},{\bm r'})$ appear in an extended 
KS equation. 
A beautiful point in this density functional theory for superconductors (SCDFT) 
is that the extended KS equation can be written in the form of the Bogoliubov-de 
Gennes equation appearing in the conventional theory for inhomogeneous 
superconductors~\cite{deGennes66}. Just as is the case with $V_{xc}({\bm r})$, 
$\Delta_{xc}({\bm r},{\bm r'})$ has no direct physical meaning, but in principle, 
if the exact form of $F_{xc}[n,\chi]$ is known, the solution of the extended 
KS equation gives us the exact result for $\chi({\bm r},{\bm r'})$, containing all 
the effects of the Coulomb repulsion including the one usually treated 
phenomenologically through the concept of $\mu^*$. As a result, we can 
determine the exact $T_c$ by the calculation of the highest temperature below 
which a nonzero solution for $\chi({\bm r},{\bm r'})$ can be found. 

In this framework of SCDFT, we can formally write down the fundamental gap 
equation to determine $T_c$ {\it exactly} as~\cite{Units-here}
\begin{eqnarray}
\Delta_j = -\sum_{j'} {\Delta_{j'} \over 
2\varepsilon_{j'}} \tanh {\varepsilon_{j'} 
\over 2T_c} \, {\cal K}_{jj'},
 \label{eq:SCDFT}
\end{eqnarray}
where $\Delta_j$ is the gap function associated with $j$th KS orbital. In just 
the same way as its energy eigenvalue $\varepsilon_j$ (which is measured 
relative to the chemical potential), $\Delta_j$ is not the quantity to be observed 
experimentally but just introduced for the mathematical convenience so as to 
obtain the exact $T_c$ by solving this BCS-type equation, Eq.~(\ref{eq:SCDFT}). 
Similarly, the pairing interaction ${\cal K}_{jj'}$, defined as the second-order 
functional derivative of $F_{xc}[n ,\chi]$ with respect to $\chi^*$ and $\chi$, 
has not any direct physical meaning, either, although this is a quantity of 
primary importance in this gap equation or even in the whole framework of SCDFT. 

Three comments are in order: (i) The functional derivatives of $F_{xc}[n ,\chi]$ 
might not be well defined, as anticipated by remembering the notorious 
energy-gap problem in semiconductors and insulators~\cite{Perdew83,Sham83,Sham85}, 
but as is ordinally the case, we shall assume that ${\cal K}_{jj'}$ is a 
well-defined quantity. (ii) In this formal derivation in SCDFT, the 
dynamical (or $\omega$-dependent) nature in the electron-electron multiple 
scatterings does not manifest itself in either the gap equation or the pairing 
interaction, in sharp contrast with the Eliashberg theory. For this reason, many 
people cast doubt on whether the physics leading to $\mu^*$ is actually taken 
into account in SCDFT. However, due to the fact that there is a very good 
correspondence between this gap equation and the one in the $G_0W_0$ approximation 
to the Elishaberg theory, as will be shown in the next section, we find that 
it is possible to include the full dynamical processes in the Cooper-pair 
formation in the framework of SCDFT, as long as the form of ${\cal K}_{jj'}$ is 
properly chosen. (iii) At $T=T_c$, ${\cal K}_{jj'}$ is evaluated at $\chi=0$. 
Thus ${\cal K}_{jj'}$ must be a functional of only the normal-state electronic 
density $n({\bm r})$. Note that each KS orbital, $j$ or $j'$, determined in the 
normal state may be regarded as a functional of $n({\bm r})$, justifying the 
view that ${\cal K}_{jj'}$ at $T=T_c$ is eventually a functional of $n({\bm r})$ 
in the normal state. 

\subsection{Application and Discussion}

This formal framework of SCDFT was not applied to actual superconductors before 
the year 2005 when an attempt was made to provide a concrete approximate form for 
$F_{xc}[n ,\chi]$ in which the contribution from the phonon-mediated attraction 
was explicitly included up to the level of the Eliashberg theory~\cite{Luders05}. 
Since then many (but mostly weakly-correlated) superconductors have been 
analyzed rather successfully in this 
framework~\cite{Sanna07,Marques05,Floris05,Profeta06,Sanna06,Floris07}. 

In the judgement of the present author, the presently available form for 
$F_{xc}[n ,\chi]$ or the one for ${\cal K}_{jj'}$ contains the information 
equivalent to that included in the Eliashberg theory for the part of the 
phonon-mediated attraction, indicating that no vertex corrections are 
considered in this treatment (amounting to the very insufficient treatment 
of the strong polaronic effect), while for the part of the Coulomb repulsion, it 
contains only very crude physics; the screening effect is treated in the Thomas-Fermi 
static-screening approximation, or the result in the random phase approximation 
(RPA) only in the static and the long-wavelength limit, neglecting both the 
dynamical and nonlocal feature in the effects of the Coulomb repulsion. This 
clearly indicates that the Coulomb repulsion is not treated on the same footing 
as the phonon-mediated attraction and this approximation for the Coulomb part 
will be just good for describing the physics represented by $\mu^*$ at usual 
metallic densities (or $r_s \approx 2$ with $r_s$ the conventional 
nondimensional density parameter) from first principles, but it fails to take 
care of the detailed dynamical nature of the screening effect, especially, 
the positive role of the plasmons in superconductivity for lower densities (or 
larger $r_s$)~\cite{Takada78,Takada93a}. Furthermore, the presently available 
form for $F_{xc}[n ,\chi]$ or ${\cal K}_{jj'}$ does not allow to discuss other 
types of the electronic mechanisms such as the spin-fluctuation one, either. In 
view of these fundamental problems, it is absolutely necessary to derive a much 
better approximate functional form for ${\cal K}_{jj'}$ for the purpose of 
investigating the electronic mechanisms in the absence/presence of the phonon 
mechanism. 

It would be appropriate here to make a rather general comment on numerical 
errors. Currently, calculations of the normal-state properties are done 
in either the local-density approximation (LDA) or the generalized gradient 
approximation (GGA)~\cite{Perdew96} to $F_{xc}[n({\bm r})]$ in DFT. We usually 
anticipate that errors in the calculated results are of the order of 1eV and 
0.3eV for LDA and GGA, respectively, and those errors are much larger than that 
expected in quantum chemistry ($\approx 0.05$eV). Now in the usual procedure in 
SCDFT, the calculation of $T_c$ (which is of the order of 0.001eV in general) is 
done simultaneously with that of the normal state and thus the error for $T_c$ 
might be of the same order as that for the normal-state properties, implying 
that it might become much larger than $T_c$ itself. 

This unfavorable situation may be avoided, if we take the following two-stage 
strategy for the calculation of $T_c$ for a family of superconducting materials 
in consideration; in the first stage, combined with available experimental 
results on the normal state, we establish a good model system representing this 
family of superconductors by making a first-principles band-structure calculation 
and then in the second stage, we evaluate $T_c$ based on the model system not 
only for reproducing the experimental $T_c$ but also for suggesting not yet 
synthesized but promising superconductors with higher $T_c$ in this family. In 
the rest of this article, we shall discuss two families of the carbon-based 
superconductors for which $T_c$s are calculated and predicted in accordance 
with this two-stage strategy. 

\section{G$_0$W$_0$ Approximation with Application to Graphite Intercalation Compounds}
\label{sec_takada.3}

\subsection{Pairing Interaction in the Weak-Coupling Region}

The three-dimensional (3D) homogeneous electron gas has been known to be a very 
useful system in constructing a successful functional form for $V_{xc}({\bm r})$ 
in either LDA or GGA by its study with use of various powerful many-body 
techniques including quantum Monte Carlo simulations. In view of this success, 
we shall study superconductivity in the same system with the conventional 
Green's-function method in order to infer a good functional form for the 
pairing interaction ${\cal K}_{jj'}$ in Eq.~(\ref{eq:SCDFT}) that will be exact 
in the weak-coupling limit. 

In a homogeneous system, momentum ${\bm p}$ is always a good quantum number and 
an electron can be specified by ${\bm p}$ and spin $\sigma$. If we write the 
electron annihilation operator by $c_{{\bm p}\sigma}$, the Hamiltonian $H$ of 
the 3D electron-gas system coupled with phonons is given by 
\begin{align}
H =& H_e+H_{ph}
\nonumber \\
=& \sum_{{\bm p}\sigma} \varepsilon_{\bm p}c_{{\bm p}\sigma}^{+}
c_{{\bm p}\sigma}
\nonumber \\
&+{1 \over 2} \sum_{{\bm q}\neq {\bm 0}}\sum_{{\bm p}\sigma}
\sum_{{\bm p'}\sigma'}V_0({\bm q})
c_{{\bm p}\sigma}^{+}c_{{\bm p'}\sigma'}^{+}c_{{\bm p'-q}\sigma}
c_{{\bm p+q}\sigma'}
\nonumber \\
&+H_{ph},
\label{eq:F0}
\end{align}
where $\varepsilon_{\bm p}\ (\equiv {\bm p}^2/2m^*\!-\!\mu)$ is the bare one-electron 
dispersion relation with $m^*$ the band mass and $\mu$ the chemical potential, 
$V_0({\bm q})\ (\equiv 4\pi e^2/\epsilon_{\infty}{\bm q}^2)$ is the bare Coulomb 
repulsion with $\epsilon_{\infty}$ the optical dielectric constant, and $H_{ph}$ 
represents all contributions containing the phonon operators. For the 
time being, there is no need of our specifying a concrete form for $H_{ph}$. 

In the thermal Green's-function method, we can treat superconductivity 
by introducing the abnormal thermal Green's function $F({\bm p},i\omega_p)$, 
which is defined at temperature $T$ by
\begin{eqnarray}
F({\bm p},i\omega_p) = -\int_0^{1/T} \! d\tau \, e^{i\omega_p \tau}
\langle T_{\tau} c_{{\bm p}\uparrow}(\tau)c_{-{\bm p}\downarrow} \rangle.
 \label{eq:F1}
\end{eqnarray}
Here $\omega_p$ is the fermion Matsubara frequency, defined by $\pi T (2p+1)$ 
with an integer $p$. At $T=T_c$ where the second-order superconducting phase 
transition occurs, this function satisfies the following formally exact gap equation: 
\begin{align}
F({\bm p},i\omega_p) \!=\!& -G({\bm p},i\omega_p)G(-{\bm p},-i\omega_p)
\nonumber \\
&\times 
T_c\! \sum_{\omega_{p'}}\!\sum_{\bm p'}\!
\tilde{J}({\bm p,p'};i\omega_p,i\omega_{p'})F({\bm p'},i\omega_{p'}),
\label{eq:F2}
\end{align}
where $G({\bm p},i\omega_p)$ is the normal thermal Green's function 
and $\tilde{J}({\bm p},{\bm p'};i\omega_p,i\omega_{p'})$ is the irreducible 
electron-electron effective interaction. 

Let us assume that the effect of interaction is weak, so that it would be enough 
to retain the terms only in lowest order in the interaction. If we adopt the 
same assumption in the calculation of the 
normal-state properties in the Green's-function approach, we are led to the 
so-called $G_0W_0$ approximation or the one-shot $GW$ approximation in 
terminology prevailing in the present-day first-principles calculation 
community, where $W$ is the effective interaction between electrons 
including both the Coulomb and the phonon-mediated interactions and $W_0$ 
represents $W$ in RPA. Incidentally, in the same kind of terminology, the 
Eliashberg theory corresponds to the $GW$ approximation. Historically, Cohen 
was the first to evaluate $T_c$ in degenerate semiconductors on the level 
of the $G_0W_0$ approximation~\cite{Cohen64,Cohen69}. Unfortunately the 
pairing interaction is not correctly derived in his theory, as explicitly 
pointed out by the present author~\cite{Takada80} who, instead, by consulting 
the pertinent work of Kirzhnits et al.~\cite{Kirzhnits73}, has succeeded in 
obtaining the correct pairing interaction~\cite{Takada78}, the result of 
which will be reiterated in the following. 

In the $G_0W_0$ approximation, we replace $G({\bm p},i\omega_p)$ by the 
bare one $G_0({\bm p},i\omega_p)\ [\equiv (i\omega_p\!-\!\varepsilon_{\bm p})^{-1}]$ 
in Eq.~(\ref{eq:F2}) and consider the case in which 
$\tilde{J}({\bm p,p'};i\omega_p,i\omega_{p'})$ is well approximated as a 
function of only the variables, ${\bm p}\!-\!{\bm p'}$ and $i\omega_p\!-
\!i\omega_{p'}$, to write 
\begin{eqnarray}
\tilde{J}({\bm p},{\bm p'};i\omega_p,i\omega_{p'})=
V({\bm p}\!-\!{\bm p'},i\omega_p\!-\!i\omega_{p'}),
\label{eq:F2a}
\end{eqnarray}
as is usually the case for $W_0$ the effective interaction in RPA, though we do 
not intend to confine ourselves to RPA at this stage. By substituting 
Eq.~(\ref{eq:F2a}) into Eq.~(\ref{eq:F2}), we obtain the gap equation in the 
$G_0W_0$ approximation as 
\begin{align}
F({\bm p},i\omega_p)\! &=\! -\!G_0({\bm p},i\omega_p)
G_0(-\!{\bm p},-\!i\omega_p)
\nonumber \\
&\times \!
T_c\! \sum_{\omega_{p'}}\!\sum_{\bm p'}\!
V({\bm p}\!-\!{\bm p'},i\omega_p\!-\!i\omega_{p'})\!F({\bm p'},i\omega_{p'}).
\label{eq:F2gap0}
\end{align}
Then, by making an analytic continuation on the $\omega$ plane to transform 
$F({\bm p},i\omega_p)$ to the retarded function $F^{R}({\bm p},\omega)$ on the 
real-$\omega$ axis and using the general relation due to the causality principle as 
\begin{eqnarray}
V^R({\bm q},\omega)=V_0({\bm q})\!
-\!\int_0^{\infty}\! \frac{d\Omega}{\pi} 
\frac{2\Omega}{\omega^2 \!-\! \Omega^2 +i\eta}
\, {\rm Im}V^R({\bm q},\Omega),
\label{eq:F2b}
\end{eqnarray}
with $\eta$ a positive infinitesimal, we end up with a gap equation for 
$F^{R}({\bm p},\omega)$. Finally, by taking the imaginary parts in both sides of 
the gap equation and integrating over the $\omega$ variable, we are led to 
an equation depending only on the momentum variable ${\bm p}$. More specifically, 
the equation can be cast into the following BCS-type gap equation: 
\begin{eqnarray}
\Delta_{\bm p} = -\sum_{\bm p'} {\Delta_{\bm p'} \over 
2\varepsilon_{\bm p'}} \tanh {\varepsilon_{\bm p'} 
\over 2T_c} \, {\cal K}_{\bm p,p'},
 \label{eq:F3}
\end{eqnarray}
where the gap function $\Delta_{\bm p}$ and the pairing interaction 
${\cal K}_{\bm p},{\bm p'}$ are, respectively, defined as 
\begin{eqnarray}
\Delta_{\bm p} \equiv 2|\varepsilon_{\bm p}|
\int_0^{\infty} \! {d\omega \over \pi}\, {\rm Im}\, F^{R}({\bm p},\omega), 
 \label{eq:F4}
\end{eqnarray}
and 
\begin{align}
{\cal K}_{\bm p},{\bm p'} &\equiv V_0({\bm p}-{\bm p'})
\!+\! \int_0^{\infty} \!\frac{2}{\pi}\,d\Omega\,
\frac{{\rm Im}V^R({\bm p}-{\bm p'},\Omega)}
{\Omega\!+\!|\varepsilon_{\bm p}|\!+\!|\varepsilon_{\bm p'}|}
\nonumber \\
&\!=\! \int_0^{\infty} \!\frac{2}{\pi}d\Omega\,
\frac{|\varepsilon_{\bm p}|\!+\!|\varepsilon_{\bm p'}|}
{\Omega^2\!+\! 
(|\varepsilon_{\bm p}|\!+\!|\varepsilon_{\bm p'}|)^2}\,
V({\bm p}\!-\!{\bm p'},i\Omega).
 \label{eq:F5}
\end{align}
With use of ${\cal K}_{\bm p},{\bm p'}$ thus derived, we can determine $T_c$ as 
an eigenvalue of Eq.~(\ref{eq:F3}), indicating that we have obtained a scheme 
in which $T_c$ is given directly from the microscopic one-electron 
dispersion relation $\varepsilon_{\bm p}$ and the effective electron-electron 
interaction $V({\bm q},i\Omega)$. Because there is no need to separate the 
phonon-mediated attraction from the Coulomb repulsion in $V({\bm q},i\Omega)$ 
and the dynamical nature of the interaction is fully taken into account, we 
can properly treat the physics leading to $\mu^*$ from first principles with 
use of this scheme. 

The very definition of the gap function $\Delta_{\bm p}$ in Eq.~(\ref{eq:F4}) 
indicates that $\Delta_{\bm p}$ does not correspond to the physical energy gap 
except in the weak-coupling limit. Similary, ${\cal K}_{\bm p},{\bm p'}$ is not a 
physical entity, although $V$ is the physical effective interaction. Both 
quantities are introduced for the mathematical convenience so as to make $T_c$ 
invariant in transforming Eq.~(\ref{eq:F2}) into Eq.~(\ref{eq:F3}). The key 
point here is that we need not solve the full gap equation~(\ref{eq:F2}) but 
much simpler one~(\ref{eq:F3}) in order to obtain $T_c$ in Eq.~(\ref{eq:F2}). 
Of course, if we want to know the physical gap function rather than 
$\Delta_{\bm p}$ to compare with experiment, we need to solve the full gap 
equation, Eq.~(\ref{eq:F2}), with $T_c$ determined by Eq.~(\ref{eq:F3}). 

Although the spin-singlet pairing has been assumed in the derivation of 
Eq.~(\ref{eq:F3}), no assumption is made on the dependence of the gap function 
on angular valuables, so that this gap equation can treat any kind of the 
pairing anisotropy in the gap function, indicating that it can be applied to 
s-wave, d-wave, $\cdots$, and even their mixture like (s+d)-wave superconductors. 

\begin{figure}[hbtp]
\begin{center}
\includegraphics[width=6.0cm]{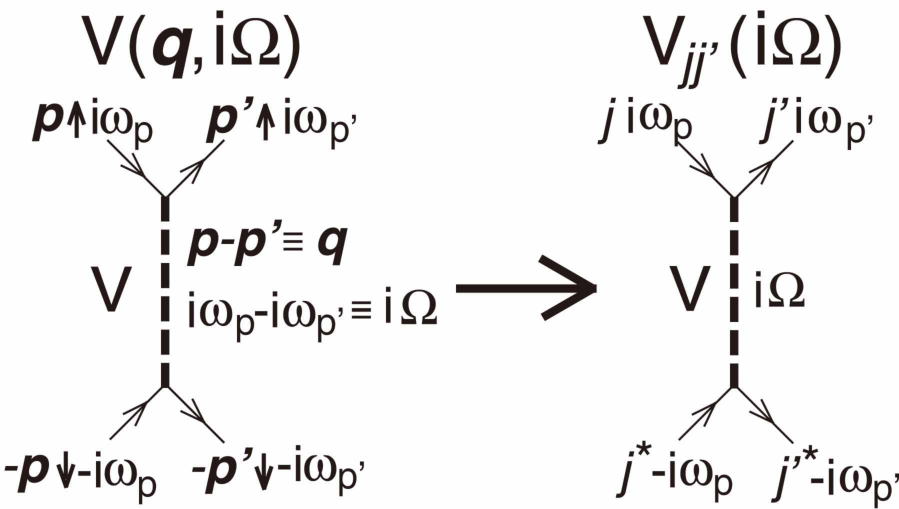}
\end{center}
\caption{Schematic representation of the dynamical effective interaction 
$V_{jj'}(i\Omega)$ as inferred from $V({\bm q},i\Omega)$.}
\label{fig1_takada}
\end{figure}

Now, let us compare Eq.~(\ref{eq:F3}) with Eq.~(\ref{eq:SCDFT}). Since the KS 
orbitals, $j$ and $j'$, can be specified by momenta, ${\bm p}$ and ${\bm p'}$, 
respectively, in a homogeneous system and $\varepsilon_{j'}=\varepsilon_{\bm p'}$, 
we may regard that these two equations are esssentially the same, suggesting 
that we may give a concrete functional form for ${\cal K}_{jj'}$ with use of 
energies of the KS orbitals, $\varepsilon_j$ and $\varepsilon_{j'}$, as
\begin{eqnarray}
{\cal K}_{jj'} = \int_0^{\infty} \!{2 \over \pi}\,d\Omega\,
{|\varepsilon_{j}|+|\varepsilon_{j'}| 
\over \Omega^2+ 
(|\varepsilon_{j}|+|\varepsilon_{j'}|)^2}\,V_{jj'}(i\Omega)\,.
 \label{eq:F5a}
\end{eqnarray}
Here $V_{jj'}(i\Omega)$ is the dynamical effective interaction working for the 
scattering process from a pair of electrons in $(j,j^*)$ orbitals to another 
pair in $(j',{j'}^*)$ orbitals, as schematically shown in Fig.~\ref{fig1_takada} 
(By $j^*$ we mean the time-reversed KS orbital of $j$.) 
We should calculate this $V_{jj'}(i\Omega)$ on the understanding that it must 
be derived from the first-principles Hamiltonian expanded with use of a 
complete set of the KS orbitals as an orthnormal basis. Together with the gap 
equation, Eq.~(\ref{eq:SCDFT}), and the KS orbitals obtained in the normal 
state by the conventional DFT-based method, Eq.~(\ref{eq:F5a}) constitutes a 
basic framework for a first-principles calculation of $T_c$ for inhomogeneous 
weak-coupling superconductors. 

\subsection{Superconductivity in Polar Semiconductors}

In order to assess the quality of this basic framework in the G$_0$W$_0$ 
approximation for calculating $T_c$ from first principles, we have applied it 
to polar degenerate semiconductors, specifically, the doped SrTiO$_3$ and 
compared the calculated results with experiments~\cite{Takada80}. 

\begin{figure}[hbtp]
\begin{center}
\includegraphics[width=6.0cm]{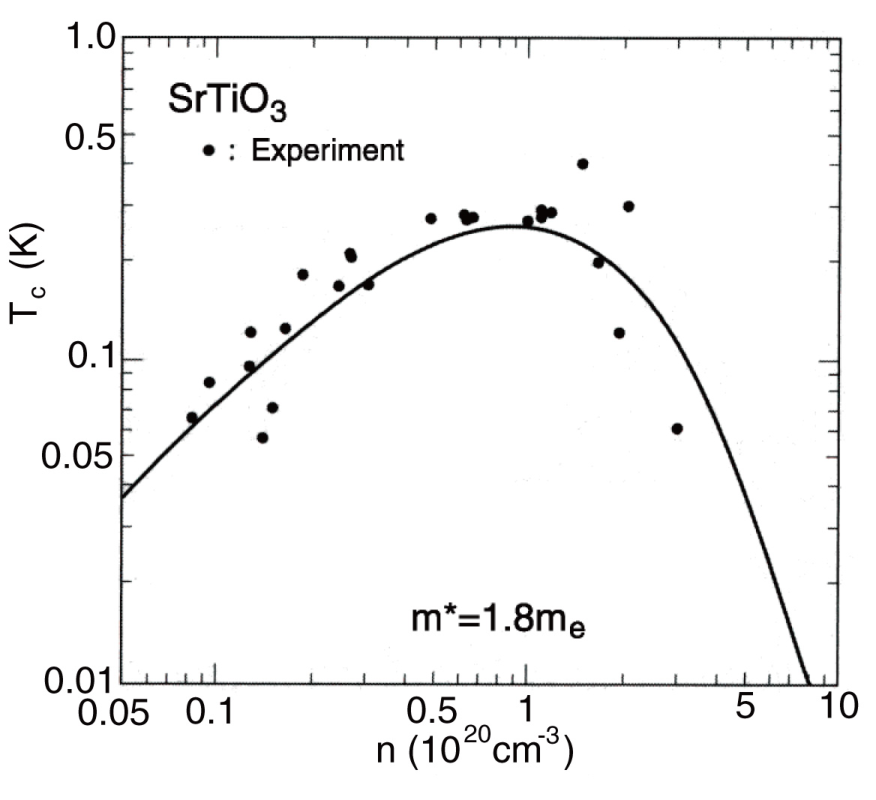}
\end{center}
\caption{Calculated results for $T_c$ in semiconducting SrTiO$_3$ 
as a function of the electron density $n$ (the solid curve ~\cite{Takada80}), 
together with the experimental results (the filled circles~\cite{Koonse}. 
See also \cite{Behnia})}
\label{fig2_takada}
\end{figure}

This material is an insulator and exhibits ferroelectricity under a uniaxial 
stress of about 0.16GPa along the [100] direction, but it turns into an $n$-type 
semiconductor by either Nb doping or oxygen deficiency, whereby the conduction 
electrons are introduced in the 3d band of Ti around the $\Gamma$ point with the 
band mass of $m^* \approx 1.8m_e$ ($m_e$: the mass of a free electron). At low 
temperatures, superconductivity appears and the observed $T_c$ shows interesting 
features; $T_c$ depends strongly on the electron concentration $n$ and it is 
optimized with $T_c \approx 0.3$K at $n \approx 10^{20}$cm$^{-3}$. Its dependence 
on the pressure is unsual; $T_c$ decreases rather rapidly with hydrostatic 
pressures, but it increases with the [100] uniaxial stress, implying that 
the superconductivity is brought about by the polar-coupling phonons associated 
with the stress-induced ferroelectric phase transition, 

Taking those situations into account, we have assumed that the material is well 
represented by a model of the 3D electron-gas system coupled with polar-optical 
phonons in which a concrete form for $V({\bm q},i\Omega)$ can be derived in RPA as 
\begin{eqnarray}
V({\bm q},i\Omega)={4\pi e^2 \over \epsilon({\bm q},i\Omega)\,{\bm q}^2 },
\label{eq:F5b}
\end{eqnarray}
with the dielectric function in the electron-optical phonon system as 
\begin{align}
\epsilon({\bm q},i\Omega)=&\epsilon_{\infty}+{4\pi e^2 \over {\bm q}^2}
\Pi_0({\bm q},i\Omega)
\nonumber \\
&+\left [ \epsilon_0({\bm q}) - \epsilon_{\infty} \right ]
{\omega_t({\bm q})^2 \over \omega_t({\bm q})^2+\Omega^2},
\label{eq:F5c}
\end{align}
where $\Pi_0({\bm q},i\Omega)$ is the polarization function in RPA (or the 
Lindhard function) for the 3D electron gas, $\omega_t({\bm q})$ is the energy 
dispersion of the transverse optical phonon, and $\epsilon_0({\bm q})$ is the 
static nonlocal dielectric function, which is determined with use of the 
static dielectric constant $\epsilon_0$ and $\omega_t({\bm q})$ as
\begin{eqnarray}
\epsilon_0({\bm q})=\epsilon_0\,{\omega_t(0)^2 \over \omega_t({\bm q})^2}. 
\label{eq:F5e}
\end{eqnarray}
The dispersionless longitudinal-phonon energy $\omega_{\ell}$ is related to 
the transverse-phonon energy in the long-wavelength limit $\omega_t(0)$ 
through the Lyddane-Sachs-Teller relation as 
\begin{eqnarray}
\omega_{\ell}= \sqrt{{\epsilon_{0} \over \epsilon_{\infty}}}\,\omega_{t}(0).
\label{eq:F5d}
\end{eqnarray}
By substituting $V({\bm q},i\Omega)$ in Eq.~(\ref{eq:F5b}) into 
Eq.~(\ref{eq:F5}) and using the experimental data to determine the values of 
parameters like $\epsilon_0$ and $\epsilon_{\infty}$ as well as the dipersion 
relation for $\omega_t({\bm q})$, we have obtained $T_c$ directly from a 
microscopic model and the results of $T_c$ are in surprizingly good quantitative 
agreement with experiment, as shown in Fig.~\ref{fig2_takada} The unusual 
dependence of $T_c$ on the pressure is also reproduced well, though it is not shown 
here. (For interested readers, refer to the original paper~\cite{Takada80}.) 
This success indicates that the present framework including the 
adoption of RPA for calculating the effective interaction is useful 
and appropriate at least in the polar-coupled phonon mechanism in which the 
contribution from the long-range part of the interaction dominates over 
that from the short-range one. 

\subsection{Graphite Intercalation Compounds (GICs)}

\subsubsection{Historical Survey}

\begin{figure}[htbp]
\begin{center}
\includegraphics[width=8.0cm]{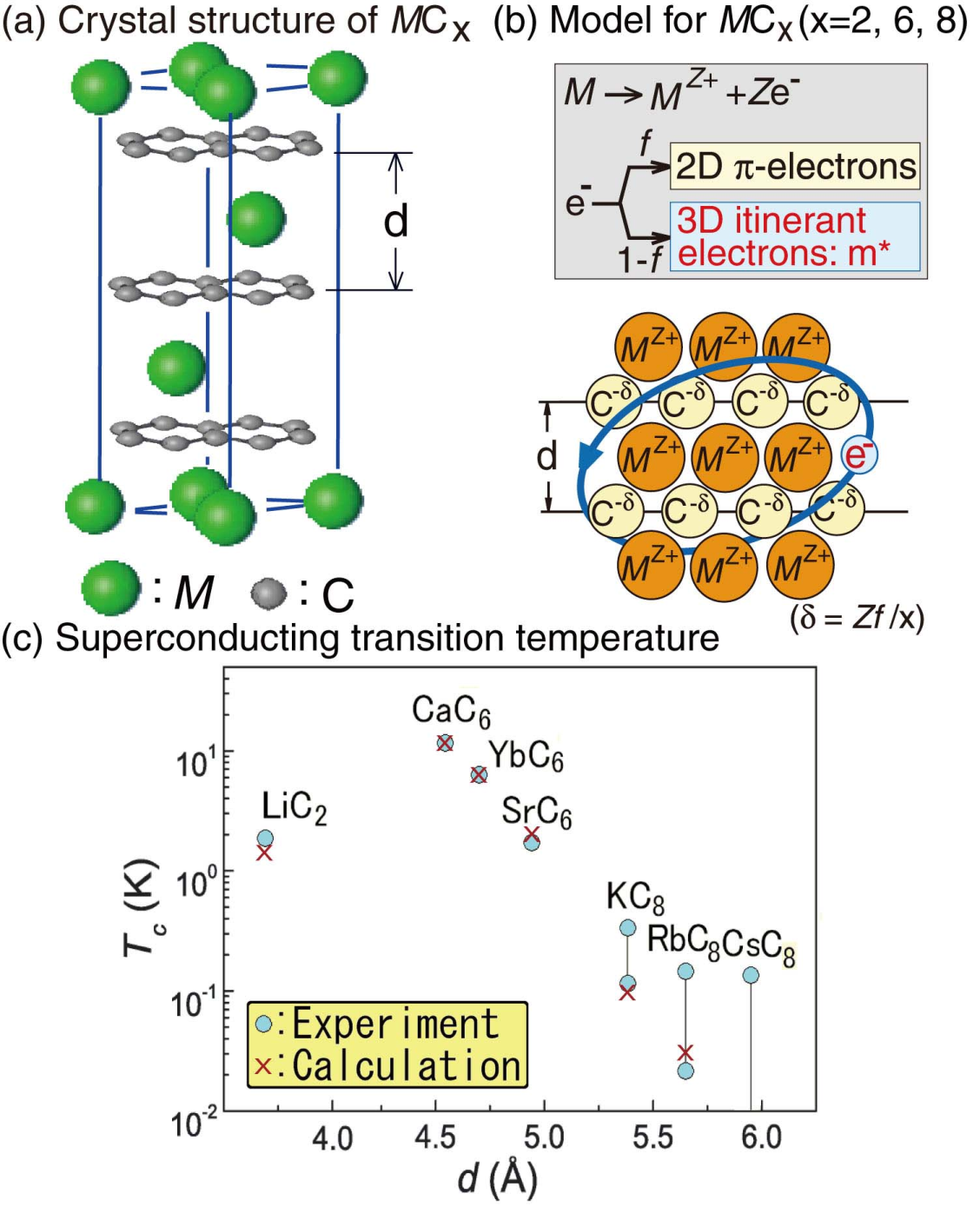}
\end{center}
\caption{(a) Crystal structure of $M$C$_x$\ ($x$=$2,\,6,\,8$). The case of $x$
=$6$ is illustrated here, in which the metal atoms, $M$s, are arranged in a rhombohedral 
structure with the $\alpha \beta \gamma$ stacking sequence, implying that 
$M$ occupies the same spot in the framework of a graphene lattice at every 
three layers (or at the distance of $3d$ with $d$ the distance between the 
adjacent graphite layers). 
(b) Simplified model to represent $M$C$_x$ superconductors. 
We consider the attraction between the 3D electrons in the interlayer band 
induced by polar-coupled charge fluctuations of the cation $M^{Z+}$ and 
the anion C$^{-\delta}$. 
(c) Superconducting transition temperature $T_c$ in the first-stage alkali- 
and alkaline-earth-intercalated graphites plotted as a function of $d$. 
The solid circles show the experimental results, while the crosses the 
calculated ones in the G$_0$W$_0$ approximation for the model represented 
in (b) with suitable values for the parameters such as $Z$, $m^*$, and $f$ 
for each $M$C$_x$.}
\label{fig3_takada}
\end{figure}

The graphite intercalation compounds (GICs) have been investigated for a long 
time from physical, chemical, and technological points of view~\cite{PhysToday1,
PhysToday2,Zabel92,AdvPhys}. Among various kinds of GICs, special attention has 
been paid to the first-stage metal compounds, mainly because superconductivity 
is observed only in this class of GICs, the chemical formula of which is written 
as $M$C$_x$, where $M$ represents either an alkali atom (such as Li, K, Rb, and 
Cs) or an alkaline-earth atom (such as Ca, Sr, and Yb) and $x$ is either $2$, 
$6$, or $8$. The crystal structure of $M$C$_x$ is shown in 
Fig.~\ref{fig3_takada}(a), in which the metal atom $M$ occupies the same spot 
in the framework of a honeycomb lattice at every $(x/2)$ layers of carbon atoms. 

The first discovery of superconductivity in GICs was made in KC$_8$ with the 
superconducting transition temperature $T_c$ of 0.15K in 1965~\cite{Hannay65}. 
In pursuit of higher $T_c$, various GICs were synthesized, mostly working with 
the alkali metals and alkali-metal amalgams as intercalants, from the late 1970s 
to the early 1990s~\cite{Koike78,Kobayashi79,Koike80,Alexander80,Iye,Belash89a,
Dresselhaus89,Belash90}, but only a limited success was achieved at that time; 
the highest attained $T_c$ was around 2-5K in the last century. For example, 
it is 1.9K in LiC$_2$~\cite{Belash89b}. 

A breakthough occurred in 2005 when $T_c$ went up to 11.5K in CaC$_6$~\cite{CaC6,
Emery} (and even to 15.4K under pressures up to 7.5GPa~\cite{Takagi}). In other 
alkaline-earth GICs, the values of $T_c$ are 6.5K and 1.65K for 
YbC$_6$~\cite{CaC6} and SrC$_6$~\cite{Kim2}, respectively. Since then, 
very intensive experimental studies have been made in those and related 
compounds~\cite{Emery,Kim2,Kim1,Kurter}. Theoretical studies have also been 
performed mainly by making state-of-the-art first-principles calculations of 
the electron-phonon coupling constant $\lambda$ to account for the observed 
value of $T_c$ for each individual superconductor~\cite{Mazin,Csanyi,Mauri,
Sanna07}. Those experimental/theoretical works have elucidated that, although 
there are some anisotropic features in the superconducting gap, the conventional 
phonon-driven mechanism to bring about s-wave superconductivity applies to those 
compounds. This picture of superconductivity is confirmed by, for example, the 
observation of the Ca isotope effect with its exponent $\alpha=0.50$, the 
typical BCS value~\cite{Hinks}. 

In spite of all those efforts and the existence of such a generally accepted 
picture, we need to know more important and fundamental issues that include: 

\noindent
(i) Standard model: Can we understand the mechanism of superconductivity in 
both alkali GICs with $T_c$ in the range $0.01-1.0$K and alkaline-earth GICs 
with $T_c$ in the range $1-10$K from a unified point of view? In other words, 
is there any standard model for superconductivity in GICs with $T_c$ ranging 
over three orders of magnitude? 

\noindent
(ii) Key parameters to control $T_c$: What is the actual reason why $T_c$ is 
enhanced so abruptly (or by about a hundred times) by just substituting K by Ca 
the atomic mass of which is almost the same as that of K? In terms of the 
standard model, what are the key controlling physical parameters to bring about 
this huge enhancement of $T_c$? This change of $T_c$ from KC$_8$ to CaC$_6$ is 
probably the most important issue in exploring superconductivity across the 
entire family of GICs. 

\noindent
(iii) Optimum $T_c$: Is there any possibility to make a further enhancement of 
$T_c$ in GICs? If possible, what is the optimum value of $T_c$ expected in the 
standard model and what kind of atoms should be intercalated to realize the 
optimum $T_c$ in actual GICs? 

Recently these three issues have been satisfactorily addressed by the present 
author~\cite{Takada09a,Takada09b}, as shall be explained one by one in the 
following three subsections. 

\subsubsection{Standard Model for Intercalated Graphite Superconductors}

The usual DFT-based self-consistent band-structure calculation is useful in 
elucidating the electronic structures of GICs in the normal state, basically 
because GICs are not strongly-correlated materials. According to such a 
calculation, no essential qualitative difference is found between alkali and 
alkaline-earth GICs. The main common features among these GICs may be 
summarized as follows: 

First of all, each intercalant metal atom in $M$C$_x$ acts as a donor and 
changes from a neutral atom $M$ to an ion $M^{Z+}$ with valence $Z$. Then, 
the valence electrons released from $M$ will transfer either to the graphite 
$\pi$ bands or the three-dimensional (3D) band composed of the intercalant 
orbitals and the graphite interlayer states~\cite{Posternak84,Holzwarth84,
Koma86}. We shall define the factor $f$ as the branching ratio between these 
two kinds of bands. Namely, $Zf$ and $Z(1-f)$ electrons will go to the $\pi$ 
and the 3D bands, respectively. The electrons in the graphite $\pi$ bands are 
characterized by the two-dimensional (2D) motion with a linear dispersion 
relation (known as a Dirac cone in the case of graphene) on the graphite layer. 

The dispersion relation of the graphite interlayer band is very similar to 
that of the 3D free-electron gas, folded into the Brillouin zone of the 
graphite~\cite{Csanyi}. Thus its energy level is very high above the Fermi 
level in the graphite, because the amplitude of the wavefunction for this band 
is small on the carbon atoms. In $M$C$_x$, on the other hand, the cation 
$M^{Z+}$ is located in the interlayer position where the amplitude of the 
wavefunctions is large, lowering the energy level of the interlayer band 
below the Fermi level. The dispersion of the interlayer band is modified 
from that of the free-electron gas because of the hybridization with the 
orbitals associated with $M$, but generally it is well approximated by 
$\varepsilon_{\bm p}={\bm p}^2/2m^*-E_{\rm F}$ with an appropriate choice of 
the effective band mass $m^*$ and the Fermi energy $E_{\rm F}$. Here the value 
of $m^*$ depends on $M$; in alkali GICs, the hybridization occurs with 
s-orbitals, allowing us to consider that $m^*=m_e$, while in alkaline-earth 
GICs, the hybridization with d-orbitals conrtibutes much, leading to 
$m^* \approx 3m_e$ in both CaC$_6$ and YbC$_6$, as revealed by 
the band-structure calculation~\cite{Mazin,Mauri}. 

The value of $f$, which determins the branching ratio $Zf:Z(1-f)$, can be 
obtained by the self-consistent band-structure calculation. In KC$_8$, for 
example, it is known that $f$ is around $0.6$~\cite{Ohno79,Wang91}. On the 
other hand, $f$ is about $0.16$ in CaC$_6$, making the electron density in 
the 3D band $n$ increase very much~\cite{Mauri}. This increase in $n$ is 
easily understood by the fact that the energy level of the interlayer band 
is much lower with Ca$^{2+}$ than with K$^+$. The concrete numbers for 
$n$ are $3.5 \times 10^{21}$cm$^{-3}$ and $2.4 \times 10^{22}$cm$^{-3}$ 
for KC$_8$ and CaC$_6$, respectively, in which the difference in both $d$ and 
$x$ is also taking into account. 

As inferred from experiments~\cite{PhysToday2,Csanyi} and also from 
the comparison of $T_c$ calculated for each band~\cite{Takada82}, 
it has been concluded that only the 3D interlayer band is responsible for 
superconductivity. Note that LiC$_6$ does not exhibit superconductivity 
because no carriers are present in the 3D interlayer band, although the 
properties of LiC$_6$ are generally very similar to those of other 
superconducting GICs in the normal state. 

With the above-mentioned common features in mind, we can think of a simple 
model of a 3D electron gas coupled with phonons for the GIC superconductors, 
which is schematically shown in Fig.~\ref{fig3_takada}(b). In order to give 
some idea about the mechanism to induce an attraction between 3D electrons in 
this model, let us imagine how each conducting 3D electron sees the charge 
distribution of the system. First of all, there are posively charged metallic 
ions $M^{Z+}$ with its density $n_M$, given by $n_M=4/3\sqrt{3}\,a^2dx$, where 
$a$ is the bond length between C atoms on the graphite layer (which is 
1.419\AA). Note that with use of this $n_M$, the density of the 3D electrons 
$n$ is given by $(1-f)Zn_M$. There are also negatively charged carbon ions 
C$^{-\delta}$ with $\delta$ given by $\delta=fZ/x$ on the average. Therefore 
the 3D electrons will feel a large electric field of the polarization wave 
coming from oscillations of $M^{Z+}$ and C$^{-\delta}$ ions created by either 
out-of-phase optical or in-phase acoustic phonons. 

Although there are some additional complications originating from the combined 
contributions from both optical and acoustic modes in the layered-lattice 
system, this coupling of an electron with the polar phonons is essentially 
similar to the one we have already considered in the previous subsection. Thus 
it is straightforward to derive the effective interaction $V({\bm q},i\Omega)$ 
in RPA in which the bare Coulomb repulsion and the polar-phonon-mediated 
attraction are treated on the same footing with the screening effects of both 
the 2D and 3D electrons. A concrete form for $V({\bm q},i\Omega)$ will not be 
given here, but for its detailed derivation we refer to the original 
paper~\cite{Takada82} in which exactly the same model as presented in 
Fig.~\ref{fig3_takada}(b) was proposed in as early as 1982 by the present 
author for analyzing superconductivity in alkali GICs. 

\subsubsection{Key Physical Parameters to Control Superconductivity}

We have evaluated $T_c$ from first principles by using $V({\bm q},i\Omega)$ 
thus obtained to solve the gap equation~(\ref{eq:F3}). 
In Fig.~\ref{fig3_takada}(c), the calculated results of $T_c$ for various 
$M$C$_x$ are plotted by the crosses with the choice of suitable values for 
the parameters such as $Z$, $m^*$, and $f$ for each material. As we see, the 
agreement between theory and experiment is quite excellent across the 
entire family of GICs, implying that our simple model may well be regarded as 
the standard one for describing the mechanism of superconductivity in GICs. 

In order to identify the controlling physical parameters to enhance $T_c$ 
in CaC$_6$ by a handred times from that in KC$_8$, let us compare the 
values for the physical parameters between the two materials: 
(i) The valence $Z$; because the valence changes from monvalence to divalence, 
the value of $Z$ in CaC$_6$ is doubled to make the bare polar-phonon-mediated 
attraction (which is in proportion to $Z^2$) stronger by four times. 
(ii) The interlayer distance $d$; it decreases from $5.42$\AA\ to $4.524$\AA, 
so that the 3D electron density $n$ increases in CaC$_6$. 
(iii) The factor $f$ to determine the branching ratio; it decreases from 
about $0.6$ to $0.16$, which also makes a further increase in $n$. 
(iv) The effective band mass for the 3D interlayer band $m^*$; it increases 
from $m_e$ to about $3m_e$, leading to a large enhancement of the density 
of states at the Fermi level.
(v) The atomic number of the ion $A$; it changes only from $39.1$ in K 
to $40.1$ in Ca. Thus the energies of phonons hardly change. 

We have recalculated $T_c$ by shifting each parameter, one by one, from the 
above-mentioned respective physical value and have found that two parameters, 
namely, $Z$ and $m^*$, are very important in controlling the overall magnitude 
of $T_c$. In fact, $T_c$ is enhanced by one order of magnitude from that in 
KC$_8$ by doubling $Z$ from $Z=1$ to $Z=2$ with $m^*$ kept to be $m_e$. 
A further enhancement of $T_c$ by another one order is seen by tripling $m^*$ 
from $m_e$ to $3m_e$, with $Z$ kept to be $Z=2$. Thus we may conclude that the 
enhancement of $T_c$ in CaC$_6$ by about a hundred times from that in KC$_8$ is 
brought about by the combined effects of doubling $Z$ and tripling $m^*$. In 
this respect, the actual value of $m^*$ is very important. Appropriateness of 
$m^* \approx 3m_e$ is confirmed not only from the band-structure 
calculations~\cite{Mazin,Mauri} but also from the measurement of the electronic 
specfic heat~\cite{Kim1} compared with the corresponding one for 
KC$_8$~\cite{Mizutani78}.

A note will be added here on the case of YbC$_6$; the basic parameters such 
as $Z$, $f$, and $m^*$ for YbC$_6$ are about the same as those for CaC$_6$, 
according to the band-structure calculation. The only big difference can be seen 
in the atomic mass; Yb (in which $A=173.0$) is much heavier than that of Ca 
by about four times, indicating weaker couplings between electrons and polar 
phonons as just in the case of comparison between KC$_8$ and RbC$_8$ or CsC$_8$. 
In fact, $T_c$ for YbC$_6$ becomes about one half of the corresponding result for 
CaC$_6$, which agrees well with experiment. One way to understand this difference 
is to regard it as an isotope effect with $\alpha \approx 0.5$~\cite{Mazin}. 

\subsubsection{Prediction of Optimum $T_c$}
\label{sec:8}

As we have seen so far, our standard model could have predicted $T_c =11.5$K 
for CaC$_6$ in 1982 and it is judged that its predictive power is very high. 
Incidentally, the author did not perform the calculation of $T_c$ for 
CaC$_6$ at that time, partly because he did not know a possibility to synthesize 
such GICs, but mostly because the calculation cost was extremely high in 
those days; a rough estimate shows that there is acceleration in computers by 
at least a millon times in the past three decades. This huge improvement 
on computational environments is surely a boost to making such a first-principles 
calculation of $T_c$ in the G$_0$W$_0$ approximation. 

\begin{figure}[htbp]
\begin{center}
\includegraphics[width=6.0cm]{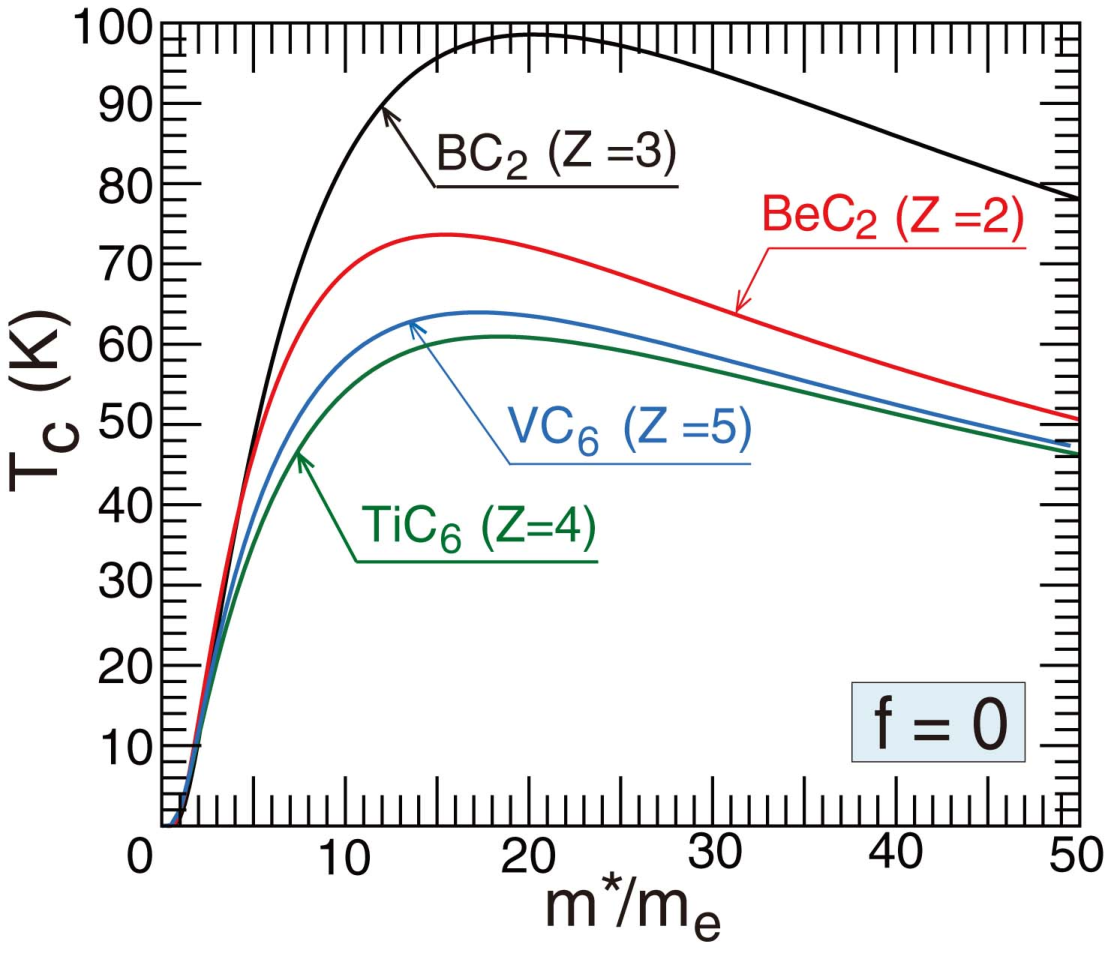}
\end{center}
\caption{Prediction of $T_c$ as a function of $m^*$ for various 
GICs in pursuit of optimum $T_c$. We assume the fractional factor $f=0$.}
\label{fig4_takada}
\end{figure}

We have explored the optimum $T_c$ in the whole family of GICs by widely 
changing various parameters involved in the microscopic Hamiltonian of 
the standard model. Examples of the calculated results of $T_c$ are shown in 
Fig.~\ref{fig4_takada}, in which $f$ is fixed to zero, the optimum 
condition to raise $T_c$, and $d$ is tentatively taken as 4.0\AA. From this 
exploration, we find that the most 
important parameter to enhance $T_c$ is $m^*$. In particular, we need $m^*$ 
larger than at least $2m_e$ to obtain $T_c$ over 10K, irrespective of any 
choice of other parameters, and $T_c$ is optimized for $m^*$ in the range 
$(10-20)m_e$. The optimized $T_c$ depends rather strongly on the parameters to 
control the polar-coupling strength such as $Z$ and the atomic mass $A$; if we 
choose a trivalent light atom such as boron to make $\omega_t(0)$ large, the 
optimum $T_c$ is about 100K, but the problem about the light atoms is that $m^*$ 
will never become heavy due to the absence of either d or f electrons. Therefore 
we do not expect that $T_c$ would become much larger than 10K, even if BeC$_2$ or 
BC$_2$ were synthesized. From this perspective, it will be much better to 
intercalate Ti or V, rather than Be or B. Taking all these points into account, 
we suggest synthesizing three-element GICs providing a heavy 3D electron system 
by the introduction of heavy atoms into a light-atom polar-crystal environment.

\section{Strong-Coupling Approach with Application to Fullerides}
\label{sec_takada.4}

\subsection{Coherence Length}

In the BCS theory for superconductivity which is applicable to superconductors 
in the weak-coupling region, $T_c$ is directly related to the coherence length 
$\xi_0$ which characterizes the spatial extent of the wave function representing 
the bound state of a Cooper pair at zero temperature. More concretely, the 
relation between them is expressed as $T_c/E_F = 2e^{\gamma}/\pi^2 (p_F\xi_0)^{-1} 
\approx 0.361/p_F\xi_0$ with $\gamma\,(=0.57721\cdots)$ the Euler-Mascheroni 
constant and $p_F$ the Fermi wave number. Because $p_F^{-1}$ is in proportion 
to the lattice constant $a_0$, the relation is rewritten into the form as 
\begin{eqnarray}
{T_c \over E_F} \ \approx \ 0.0735\,{a_0 \over \xi_0}
\label{eq:F15}
\end{eqnarray}
for a monovalent metal in the fcc-lattice structure. (For other valence and/or 
crystal structure, the coefficient of 0.0735 changes, but it always remains in 
the same order of magnitude.) For usual elemental superconductors, $T_c/E_F$ is 
of the order of $10^{-4}$ and thus $\xi_0$ is about a thousand times larger than 
$a_0$, validating the approach in momentum space. On the other hand, the relation 
in Eq.~(\ref{eq:F15}) implies that high $T_c$ is inevitably associated with short 
$\xi_0$. In fact, $\xi_0$ is observed as only a few nm or less, i.e., of the 
same order of $a_0$ in many of the recently synthesized high-$T_c$ 
superconductors~\cite{Text} such as the cuprates, the alkali-doped C$_{60}$, 
and MgB$_2$. 

A similar message can be obtained through the so-called Uemura 
plot~\cite{Uemura91a,Uemura91b,Takada92}, according to which there is a universal 
relation of $T_c/E_F \approx 0.04$ for a wide variety of strong-coupling 
superconductors. If this relation is put into Eq.~(\ref{eq:F15}), we obtain 
$\xi_0 \approx 2\,a_0$. Furthermore, if we think of the Bose-Einstein 
condensation (BEC) for an assembly of very tightly-bound pairs of electrons, 
the condensation temperature (which amounts to $T_c$ in such a system) is given 
by $T_c/E_F=(2/9\pi\, \zeta(3/2)^2)^{1/3}\approx 0.218$, where $\zeta(3/2)\,
(=2.6124\cdots)$ is the Riemann's zeta function $\zeta(x)$ at $x=3/2$. This 
value of $T_c/E_F$ (which must be the optimum value for fermionic superconductors) 
suggests $\xi_0 \approx 0.3\,a_0$. Thus, in treating superconductivity in the 
strong-coupling region, we need to consider a situation of extremely short $\xi_0$, 
validating the approach in real space, which is totally different from that of 
very long $\xi_0$ in the weak-coupling superconductors. 

\subsection{Pairing Interaction in the Strong-Coupling Region}

In view of the above-mentioned difference in $\xi_0$, we shall exploit the 
shortness of $\xi_0$ in reformulating the problem of making a quantitative 
calculation of $T_c$ for strong-coupling superconductors in the Green's-function 
approach~\cite{Takada07}. Let us start with this reformulation by considering 
the dynamical correlation function for singlet pairing of two electrons in the 
normal state $Q_{sc}^R({\bm q},\omega)$, which is defined as 
\begin{eqnarray}
Q_{sc}^R({\bm q},\omega)=-i \int_0^{\infty}dt\,e^{i\omega t -\eta t} 
\langle[e^{iHt}\Phi_{\bm q}e^{-iHt},\Phi_{\bm q}^+]\rangle,
\label{eq:F16}
\end{eqnarray}
where $H$ is the Hamiltonian for a homogeneous electron system and $\Phi_{\bm q}$ 
is the electron-pair annihilation operator, defined by $\Phi_{\bm q}\,\equiv 
\sum_{\bm p}c_{{\bm p}+{\bm q}\uparrow}c_{-{\bm p}\downarrow}$. In terms of the retarded 
pairing correlation function $Q_{sc}^R({\bm q},\omega)$, we can define $T_c$ as 
the temperature at which $Q_{sc}^R({\bm q},\omega)$ diverges at ${\bm q} = {\bm 0}$ 
in the static limit ($\omega \to 0$) with the decrease of $T$. 

\begin{figure}[htbp]
\begin{center}
\includegraphics[width=6.5cm]{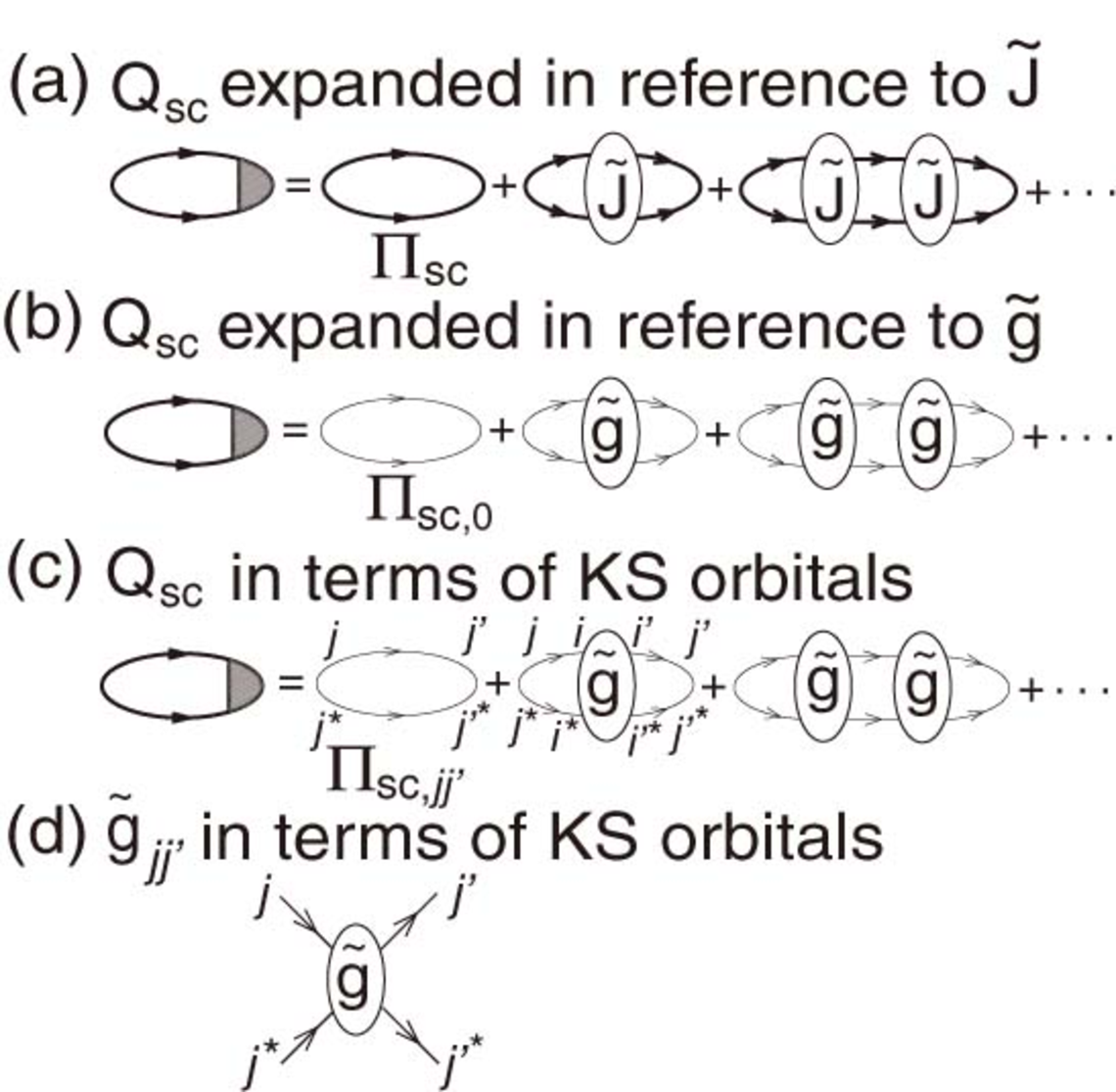}
\end{center}
\caption{Pair correlation function $Q_{sc}$ described in terms of using either 
(a) the pairing polarization function $\Pi_{sc}$ and the irreducible 
electron-electron effective interaction $\tilde{J}$ or (b) the noninteracting 
pairing polarization function $\Pi_{sc,0}$ and the effective interaction 
${\tilde g}$. By the extension of the idea of expanding $Q_{sc}$ with use of 
$\Pi_{sc,0}$, we show a scheme to express it in terms of Kohn-Sham 
orbitals in (c), together with the schematic representation of ${\tilde g}_{jj'}$ 
in (d).}
\label{fig5_takada}
\end{figure}

As shown schematically in Fig.~\ref{fig5_takada}(a), $Q_{sc}$ is conventionally 
divided into a sum of terms classified by the number of $\tilde{J}$s, where 
$\tilde{J}$, which has already appeared in Eq.~(\ref{eq:F2}), is the irreducible 
electron-electron effective interaction including all vertex corrections. Using 
$\Pi_{sc}({\bm q},\omega)$ the pairing polarization function composed of two 
{\it full} electron Green's functions including all self-energy corrections, we 
can express the infinite sum in Fig.~\ref{fig5_takada}(a) in a quite symbolical 
way as 
\begin{eqnarray}
Q_{sc}=-{\Pi_{sc} \over 1+{\tilde J}\,\Pi_{sc}}.
\label{eq:F17}
\end{eqnarray}
Then, the divergence in $Q_{sc}^R({\bm 0},\omega \to 0)$ occurs at the zero of 
the denominator in Eq.~(\ref{eq:F17}), which provides exactly the same $T_c$ as 
that obtained through the solution of Eq.~(\ref{eq:F2}). 

Now, instead of using $\Pi_{sc}$, we shall consider an alternative expansion 
with use of $\Pi_{sc,0}$ the pairing polarization function composed of two 
{\it bare} electron Green's functions. As shown schematically in 
Fig.~\ref{fig5_takada}(b), if we introduce the pairing interaction ${\tilde g}$ 
by the definition of 
\begin{eqnarray}
{\tilde g} \equiv {\tilde J}+{1 \over \Pi_{sc}}-{1 \over \Pi_{sc,0}},
\label{eq:F19}
\end{eqnarray}
we can rewrite $Q_{sc}$ in Eq.~(\ref{eq:F17}) into another exact form as 
\begin{eqnarray}
Q_{sc}=-{\Pi_{sc,0} \over 1+{\tilde g}\,\Pi_{sc,0}}.
\label{eq:F18}
\end{eqnarray}
Since we can always calculate $\Pi_{sc,0}$ easily, the problem of estimating 
$T_c$ is reduced to the evaluation of ${\tilde g}$ in the limit of $\omega \to
 0$ at ${\bm q}={\bm 0}$. Note that ${\tilde g}$ does not depend strongly on either 
${\bm q}$ or $\omega$, in sharp contrast with $\Pi_{sc,0}^{R}({\bm q},\omega)$. 
In the BCS theory, for example, ${\tilde g}$ is taken as a constant which 
represents a weakly attractive and spatially local interaction working only 
in the vicinity of the Fermi level, given that the phonon energy $\omega_0$ 
is much smaller than $E_F$. Then the zero of the denominator in Eq.~(\ref{eq:F18}) 
at ${\bm q}={\bm 0}$ and $\omega \to 0$ provides the well-known BCS formula for $T_c$. 
Note also that the problem of finding the zero of $1+{\tilde g}\,
\Pi_{sc,0}^R({\bm 0},\omega \to 0)$ is just the same as that of solving the 
gap equation in Eq.~(\ref{eq:F2gap0}) with replacing $V({\bm p}\!-\!{\bm p'},
i\omega_p\!-\!i\omega_{p'})$ by ${\tilde g}$. 

In the case of short $\xi_0$, we may assume that the essential physics of 
electron pairing can be captured even if we treat only a small-cluster system, 
as long as the system size is large enough in comparison with $\xi_0$. Under this 
assumption, we can take the following procedure for determining ${\tilde g}$: 
First, by representing the values for $\Pi_{sc,0}$ and ${\tilde g}$ in an 
$N$-site system as $\Pi_{sc,0}^{(N)}$ and ${\tilde g}_N$, respectively, we may 
write the pairing correlation function for the $N$-site system $Q_{sc}^{(N)}$ as 
\begin{eqnarray}
Q_{sc}^{(N)}=-\,{\Pi_{sc,0}^{(N)} \over 1+{\tilde g}_N\Pi_{sc,0}^{(N)}},
\label{eq:F20}
\end{eqnarray}
in accordance with Eq.~(\ref{eq:F18}). In general, the exact bulk value 
${\tilde g}$ will be obtained by taking the large-$N$ limit of ${\tilde g}_N$, 
but expecting that ${\tilde g}_{N_0}$ with a small positive integer 
$N_0$ (for example, $N_0 \approx 2$) is already close to ${\tilde g}$, we may 
estimate ${\tilde g}$ by the relation of 
\begin{eqnarray}
{\tilde g} = -{1 \over Q_{sc,0}^{(N_0)}}-{1 \over \Pi_{sc,0}^{(N_0)}},
\label{eq:F21}
\end{eqnarray}
with $Q_{sc}^{(N_0)}$ evaluated rigorously, for example, by exact diagonalization. 
Of course, we can determine a better value of ${\tilde g}$ by checking the 
saturation behavior of $\{{\tilde g}_{N_0}\}$ with the increase of $N_0$. 

\subsection{Interpolation Formula for the Pairing-Interaction Functional}

Based on the knowledge so far obtained, let us speculate an appropriate 
functional form for the pairing interaction ${\cal K}_{jj'}$ in SCDFT. 
It would be natural to expand the electron field operator in terms of the 
Kohn-Sham orbitals $\{|j\rangle\}$ in an inhomogeneous electron system, so that 
we can define the electron-pair operator $\Phi_j$ as $\Phi_{j}\,\equiv 
c_{j}c_{{j}^*}$ with $|{j}^*\rangle$ the time-reversed orbital of $|{j}\rangle$. 
Then, we can introduce the pairing polarization function in terms of $\Phi_j$ 
and $\Phi_{j'}^+$. In particular, we can calculate this quantity in the 
noninteractiong system $\Pi_{sc,jj'}$ in analogy with $\Pi_{sc,0}$, as 
schematically introduced in Fig.~\ref{fig5_takada}(c). At the same time, we can 
define a quantity ${\tilde g}_{jj'}$ as schematically shown in 
Fig.~\ref{fig5_takada}(d). 

In the strong-coupling region, we do not expect that this ${\tilde g}_{jj'}$ 
exhibits strong dependence on $i\Omega$, because the Fermi energy, the scale 
for the electron kinetic energy, must be much smaller than the energy scale 
representing electron-electron interactions. In addition, ${\tilde g}_{jj'}$ 
does not depend strongly on orbital variables, either, because the pairing 
interaction should be short-ranged in real space in line with the shortness of 
$\xi_0$. Thus we may regard ${\tilde g}_{jj'}$ as a constant. Then, if we take 
${\cal K}_{jj'}$ in Eq.~(\ref{eq:SCDFT}) as ${\tilde g}_{jj'}$, we can easily 
see that both Eq.~(\ref{eq:SCDFT}) and the zero of $1+{\tilde g}_{jj'}\,
\Pi_{sc,jj'}^R$ provide the same $T_c$. 

In a more general situation of the intermediate-coupling region, however, 
${\tilde g}_{jj'}$ will depend on both $i\Omega$ and orbital variables. In order 
to treat such a situation and also by paying attention to the weak-coupling 
situation examined in the previous section, we propose the functional form for 
${\cal K}_{jj'}$ in exactly the same form as that in Eq.~(\ref{eq:F5a}) with 
replacing $V_{jj'}(i\Omega)$ by ${\tilde g}_{jj'}(i\Omega)$. Note that this 
functional form of ${\cal K}_{jj'}$ is reduced to ${\tilde g}_{jj'}(0)$, if 
${\tilde g}_{jj'}(i\Omega)$ does not depend on $\Omega$, assuring us that 
the framework in SCDFT with this choice of ${\cal K}_{jj'}$ provides 
the same $T_c$ as that in the usual Green's-function approach in the 
strong-coupling region. This means that we can successfully take care of both 
self-energy and vertex corrections beyond the G$_0$W$_0$ approximation by 
upgrading $V_{jj'}(i\Omega)$ to ${\tilde g}_{jj'}(i\Omega)$ in Eq.~(\ref{eq:F5a}).

\subsection{Alkali-Doped Fullerides}

\subsubsection{Aims of This Subsection}

The fulleride is a molecular crystal with narrow threefold conduction bands 
(with the band width $W \approx 0.5$eV) derived from the $t_{1u}$ electronic 
levels of each C$_{60}$ molecule. With the doping of three alkali atoms per one 
C$_{60}$ molecule, we obtain the half-filled situation in which superconductivity 
occurs with $T_c$ in the range $18-38$K~\cite{Hebard91,Takabayashi09}and the 
short coherence length $\xi_0$ of only a few molecular units. As for the 
mechanism of superconductivity in the alkali-doped fullerides, phonons are 
widely believed to play an important role. This belief is based upon a crude 
estimate of $T_c$ in the conventional Eliashberg theory. In fact, analysis of 
various experiments with use of this theory has shown that many aspects of 
superconductivity in these fullerides are consistent with a picture of $s$-wave 
BCS superconductors with the Cooper pairs driven by the coupling to the 
intramolecular high-frequency phonons in $H_g$ symmetry (with the phonon energy 
$\omega_0 \approx 0.2$eV and the nondimensional electron-phonon coupling constant 
$\lambda \approx 0.6$ for Rb$_3$C$_{60}$)~\cite{Gunnarsson97,Ramirez,Gelfand}. 

The above picture, however, seems to be too much simplified and we may raise 
several fundamental questions. From a theoretical point of view, one 
of the serious problems is ill foundation to adopt the Eliashberg theory in the 
fullerides due to the importance of vertex corrections~\cite{Takada93b,vertex}. 
From an experimental side, the following four experimental facts have been 
observed which we cannot easily understand with use of the Eliashberg theory: 
(i) The relation between $T_c$ and the lattice constant $a_0$ changes remarkably 
when the crystal structure changes from the face centered cubic (fcc) to simple 
cubic (sc) lattice by the introduction of Na, a smaller ion compared to K, Rb, 
or Cs, as a dopant ion~\cite{Tanigaki}. (ii) The antiferromagnetic (AF) 
insulating behavior has been reported in ammoniated K$_3$C$_{60}$, which is 
peculiar in the sense that $s$-wave BCS superconductivity exists in the 
vicinity of an AF phase~\cite{Iwasa}. A similar problem is also seen in the 
body-centered cubic A15-structured Cs$_3$C$_{60}$~\cite{Takabayashi09}. 
(iii) The anomalous $^{13}$C isotope effect on $T_c$ for 50\% $^{13}$C 
substitution has been observed, reflecting the difference between the molecular 
and atomic mixture of $^{12}$C and $^{13}$C atoms~\cite{Lieber}. (iv) With the 
deviation of electron number per site $n$ from half-filling, $T_c$ decreases 
rapidly in both sides of the deviation~\cite{Yildirim}. 

Although the experiment (i) may be reproduced in the Eliashberg theory with some 
judicious choice of parameters, the rest (ii)-(iv) cannot be explained even 
qualitatively in the theory. In addition, the behavior of $T_c$ as a function 
of $n$ observed in the experiment (iv) cannot be predicted either by those 
theories proposed so far to account for the copper-oxide high-$T_c$ 
superconductivity based on some strong-correlation models, though such models 
may be favorable for the explanation of the experiment (ii). 

A successful theory for the fullerene superconductors should not only reproduce 
these experiments in a coherent fashion but also clarify the reason why the 
effect of vertex corrections, even if it is large, does not manifest itself 
in many superconducting properties. In quest of such a theory from a somewhat 
general viewpoint, the present author made intensive studies of the 
Hubbard-Holstein (HH) model and its extension in the past and successfully 
explained all the issues (i)-(iv) raised above~\cite{Takada98,Takada1,Takada2,
Takada3,Hotta1,Hotta}. In the rest of this subsection, we shall focus on the 
first three issues in order to show how nicely the HH model applies to the 
fullerides, as far as $T_c$ is concerned. Then we shall explore a possibility 
to enhance $T_c$ in this class of superconductors by changing the parameters 
involved in the model. 

\subsubsection{The Hubbard-Holstein Model}

In a molecular crystal, it is a very good approximation to regard each molecular 
unit as a ``site'' in a lattice. We shall adopt this approximation and describe 
the electron-phonon system in the alkali-doped fullerides by a model Hamiltonian 
$H$ in site representation in which each site corresponds to each C$_{60}$ 
molecular unit. Since the conduction band width is narrow, the intermolecular 
hopping integral $t$ must be small, indicating that only the nearest-neighbor 
hopping is relevant in modelling the kinetic energy of the system. Then, it 
would be appropriate to decompose $H$ into the nearest-neighbor electron-transfer 
term $H_t$ and the sum of site terms $\sum_i H_i$ including both the 
electron-electron and the electron-local phonon interactions. In order to 
faithfully represent the threefold degenerate $t_{1u}$-conduction bands coupled 
with eight $H_g$ intramolecular Jahn-Teller (JT) phonons, we would need to 
include the $t_{1u} \otimes H_g$ JT structure in $H_i$, as has often been the 
case~\cite{Varma,Manini94a,Manini94b,Gunnarsson03}. In making a detailed study 
on the stability of the AF insulating phase, it is known that this feature of 
band multiplicity plays a rather crucial role~\cite{Gunnarsson00}, but in 
treating superconductivity itself, the band multiplicity is not considered to 
be of primary importance~\cite{Cappelluti05}. Therefore we shall take a simplest 
possible model, namely, the {\it one-band} HH model to discuss superconductivity 
from a more general viewpoint that will be applicable to the whole family of 
fullerene superconductors including not only electron-doped but also hole-doped 
materials with different JT structures. 

The concrete form for $H$ in the HH model is given by 
\begin{eqnarray}
  H = - t \sum_{ \langle i,i' \rangle\,\sigma}
  (c_{i\sigma}^{\dagger}c_{i'\sigma}+{\rm h.c.})+\sum_i H_i,
\label{HHmodel-1}
\end{eqnarray}
with the site term $H_i$, written as
\begin{align}
H_i&=   - \mu \sum_{\sigma} n_{i\sigma} + U\, n_{i\uparrow}n_{i\downarrow}
\nonumber \\
&+\sqrt{\alpha}\omega_0 \sum_{\sigma}n_{i\sigma}(a_i+a_i^{\dagger})
+\omega_0\, a_i^{\dagger}a_i,
\label{HHmodel-2}
\end{align}
where $\langle i,i' \rangle$ represents the nearest-neighbor-site pair, 
$c_{i\sigma}$ the operator to annihilate a spin-$\sigma$ electron at site $i$, 
$\mu$ the chemical potential, $n_{i\sigma} \equiv c_{i\sigma}^{\dagger}
c_{i\sigma}$, $U$ the on-site (or intramolecular) Coulomb repulsion, 
$\alpha$ the nondimensional electron-phonon coupling constant which is related 
to the conventional electron-phonon coupling constant $\lambda$ through 
$\lambda=\alpha \omega_0 /tz$ with $\omega_0$ the optical-phonon energy and 
$z$ the coordination number, and $a_i$ the operator to annihilate an optical 
phonon at site $i$.

The characteristic features of the fullerenes can be well captured by an 
appropriate choice of the parameters involved in this Hamiltonian. In fact, 
the narrowness of the conduction band can be described by the smallness of $t$ 
of the order of 0.1eV. The difference in the crystal structure as well as the 
effect of band multiplicity can be well accounted for by a suitable choice of $z$. 
The high-frequency intramolecular optical phonons coupled strongly to electrons 
can be treated by considering the local phonons with the energy $\omega_0\, 
(\approx 0.2$eV) at each site. The short-range Coulomb potential must be 
relevant in the fullerenes due to the proximity to the AF state and it can be 
included by the introduction of $U$. Note that this $U$ is not the 
direct Coulomb repulsion between electrons on a carbon atom $U_{\rm atom}$ 
which is of the order of 5-10eV. Rather it is the sum of $U_{\rm atom}$ and 
the attraction $-U_{\rm pol}$ due to the electronic polarization effect of 60 
$\pi$-electrons in the C$_{60}$ molecule~\cite{Chakravarty}. Because of a strong 
cancellation between $U_{\rm atom}$ and $-U_{\rm pol}$, $U$ is expected to 
be of the order of 0.1eV, which is about the same magnitude as that for the 
phonon-mediated attraction, $-U_{ph} \equiv -2\alpha\omega_0$. 

Intensive studies on the ground state in this system by exact diagonalization 
of small-size clusters have revealed that the half-filled HH model exhibits 
interesting competition among charge-density-wave (CDW), spin-density-wave (SDW), 
and superconducting states~\cite{Takada2,Hotta1}. We may summarize the results 
in the following way: (a) If $U-U_{ph}$ is at least smaller than $-t$, the CDW 
state composed of an array of immobile bipolarons is stabilized. (b) If 
$U-U_{ph}$ is larger than $t$, there appears the SDW state which is nothing but 
the AF state in this case. (c) Superconductivity appears only in the CDW-SDW 
transition region where $|U-U_{ph}|$ is less than $t$. In this offset situation 
of $U \approx U_{ph}$, the state may be regarded as an assembly of nearly free 
polarons and the main effect of the strong electron-phonon vertex corrections 
is found to form a polaron from a bare electron. This implies that the 
Eliashberg theory or even the BCS theory is expected to be accurate enough to 
describe superconductivity in this system, if it is applied on the basis of the 
polaron picture~\cite{Takada2,Hotta1,Alexandrov1}. In any case, the intrisic 
competition of superconductivity with the AF state in the half-filled HH model 
resolves the issue (ii) raised previously.

\begin{figure}[htbp]
\begin{center}
\includegraphics[width=7.0cm]{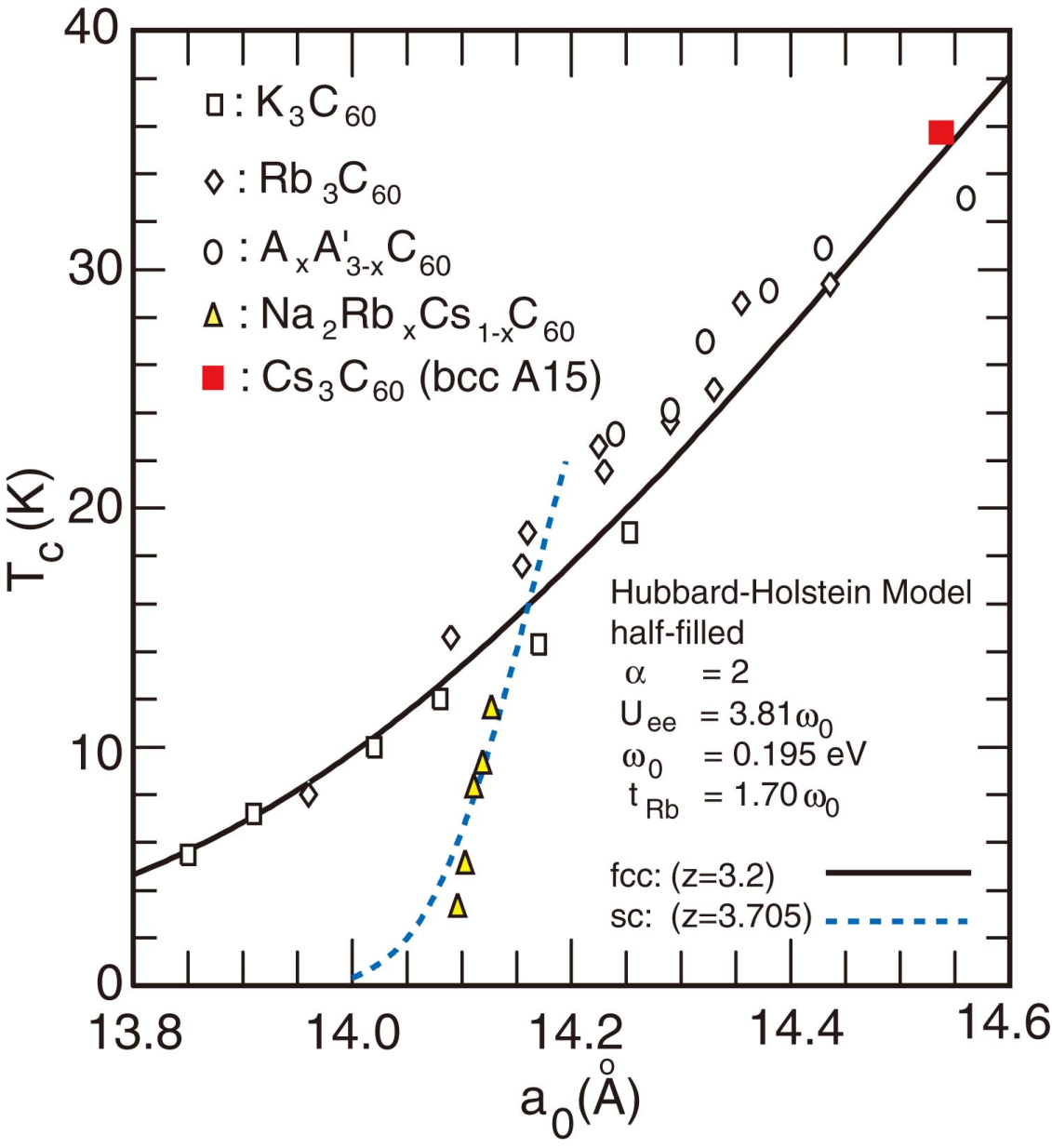}
\end{center}
\caption{$T_c$ as a function of the lattice constant $a_0$ in the
alkali-doped fullerides and its dependence on the crystal structure.
The solid and the broken curves represent our theoretical results, while the 
experimental results are given by square, triangle, diamond, and circle marks.}
\label{fig6_takada}
\end{figure}

In order to address the issue (i), we have applied our framework to calculate 
$T_c$ to the HH model by estimating ${\tilde g}$ with use of a cluster as small 
as two molecules (i.e., $N_0=2$). We have evaluated the variation of $t$ with 
the change of the lattice constant $a_0$ by fitting the change in the band width 
obtained by the band-structure calculation for each lattice 
structure~\cite{Satpathy,Saito}. More specifically, $t$ is determined by
\begin{eqnarray}
t = t_{\rm Rb}\,{d \over d_{\rm Rb}}\,\exp \left (-{d-d_{\rm Rb} \over \Delta} 
\right),
\label{t-1}
\end{eqnarray}
with $d=a_0/\sqrt{2}-6.95$\AA\,, $\Delta=0.55$\AA\,, and the corresponding 
parameters, $t_{\rm Rb}$ and $d_{\rm Rb}$, for Rb$_3$C$_{60}$. By choosing 
$t_{\rm Rb}=1.70\omega_0$ suitable for Rb$_3$C$_{60}$ and the common parameters 
such as $U=3.81\omega_0$, $\alpha=2$, and $\omega_0=0.195$eV with reference to 
band-structure calculations and relevant experiments, we obtain the results 
for $T_c$ as a function of $a_0$, which agrees remarkably well with 
experiment as clearly demonstrated in Fig.~\ref{fig6_takada} Note that in our 
calculation, the difference in $T_c$ between fcc and sc structures arises mainly 
from that in the ``effective'' lattice coordination number $z$~\cite{Gelfand}. 
This indicates that $z$ is a key to the resolution of the issue (i). Note also 
that the recent experimental result for Cs$_3$C$_{60}$ under pressure is also 
on our theoretical curve, if we estimate an appropriate value of $a_0$ so as 
to reproduce the same volume per one C$_{60}$ molecule for this bcc A15 structure. 

The competing feature between $U_{ph}$ and $U$ is also identified to be 
a key to the resolution of the issue (iii)~\cite{Takada3}. The observed 
anomalous isotope effect cannot be explained by either the phonon or the 
electronic polarization mechanism alone. It can be reproduced, if both these 
mechanisms are included simultaneously. Namely, we need to consider the change 
in both $\omega_0$ and $U$ induced by the isotope substitution. The sensitivity 
of $T_c$ to the local change in the Coulomb potential $U$ is due to the 
very short-range nature of $\xi_0$ of this superconductivity. 

\subsubsection{Prospect for Room-Temperature Superconductors}

\begin{figure}[htbp]
\begin{center}
\includegraphics[width=7.2cm]{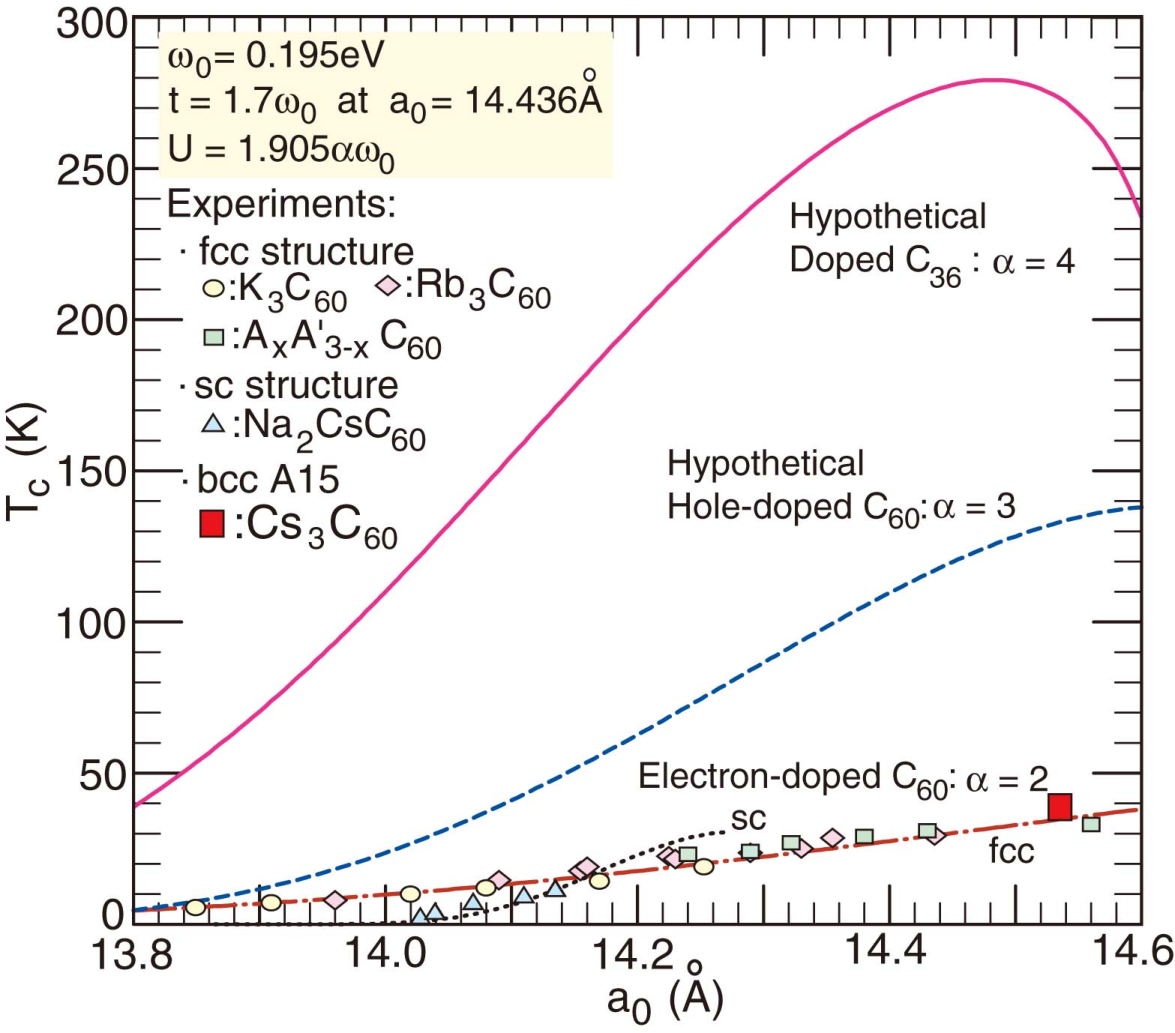}
\end{center}
\caption{$T_c$ as a function of the lattice constant $a_0$. For electron-doped 
C$_{60}$ with $\alpha=2$, the theoretical curves are drawn, together with the 
experimental results. We also give the results for both $\alpha=3$ and $4$ 
corresponding to (hypothetical) hole-doped C$_{60}$ and doped C$_{36}$.}
\label{fig7_takada}
\end{figure}

Encouraged by the success in reproducing $T_c$ for the alkali-doped fullerene 
superconductors by the application of our theoretical framework to the HH model 
at half-filling in the nearly offset situation of $U \approx U_{ph}$, we have 
attempted to explore the optimum value of $T_c$ with the increase of the 
electron-phonon coupling constant $\alpha$, whereby $U$ is also increased to 
keep the offset situation. Examples of the calculated results of $T_c$ are 
plotted in Fig.~\ref{fig7_takada} As is seen, $T_c$ goes beyond 100K for 
$\alpha=3$ and it reaches room temperature for $\alpha=4$. 

By the band-structure calculation~\cite{Mazin3}, it is known that $\alpha$ is 
about three for the hole-doped C$_{60}$ in which hole carriers will be introduced 
into the $h_u$-valence bands. On the other hand, in the crystal composed of 
C$_{36}$, $\alpha$ will be about four~\cite{Cohen}, if the C$_{36}$ solid is 
successfully doped to have an enough number of mobile carriers in the system, 
although it seems to be a very difficult task~\cite{Zettl}. 

\section{Conclusion and Discussion}
\label{sec_takada.5}

In this chapter, we have proposed a new functional form for the pairing 
interaction ${\cal K}_{jj'}$, a key quantity in the gap equation, 
Eq.~(\ref{eq:SCDFT}), to determine $T_c$ in the density functional theory 
for superconductors. The functional form is given in Eq.~(\ref{eq:F5a}) 
with the effective electron-electron interaction $V_{jj'}(i\Omega)$ replaced 
by ${\tilde g}_{jj'}(i\Omega)$ defined schematically in 
Fig.~\ref{fig5_takada}(d). We have assessed the usefulness of this functional 
form by its applications to both the weak-coupling superconductors like the 
alkali- and alkaline-earth-intercalated graphites and the strong-coupling 
superconductors like the alkali-doped fullerides. We have also explored a 
possibility to enhance $T_c$ up to room temperature. 

The proposed functional form has just opened a challenging frontier of the 
research on high-$T_c$ superconductivity. Although we need to improve on the 
functional form itself by utilizing the information obtained by its application 
to a much wider range of materials, we can think of several ways to make use of 
this new theoretical tool. For example, we can make a quantitative assessment 
of the effectiveness of each superconducting mechanism, either phononic or 
electronic, so far proposed by investigating how much $T_c$ is actually enhanced 
with the introduction of the mechanism. The knowledge accumulated by such 
investigations will pave the way to reach a room-temperature superconductor. 

\section{Acknowledgments}
This work is supported by a Grant-in-Aid for Scientific Research (C) 
(No. 21540353) from MEXT, Japan. 




\begin{references}
\bibitem{Units-here} Throughout this article, we use the units in which 
$\hbar = k_{\rm B} = 1$.
\end{references}
\end{document}